\documentclass[aps,showpacs,preprintnumbers,nofootinbibt,twocolumn]{revtex4}

\usepackage{amssymb}
\usepackage{graphicx}
\usepackage{color}

\def\be{\begin{equation}}
\def\ee{\end{equation}}
\def\bea{\begin{eqnarray}}
\def\eea{\end{eqnarray}}

\begin{document}

\title{Can Superconducting Cosmic Strings Piercing Seed Black Holes Generate Supermassive Black Holes in the Early Universe?}
\author{Matthew J. Lake$^{1,2}$}
\email{matthewj@nu.ac.th}
\author{Tiberiu Harko$^{3,4}$}
\email{t.harko@ucl.ac.uk}
\affiliation{$^1$ The Institute for Fundamental Study, ``The Tah Poe Academia Institute",
\\ Naresuan University, Phitsanulok 65000, Thailand and \\
$^2$ Thailand Center of Excellence in Physics, Ministry of Education,
Bangkok 10400, Thailand }
\affiliation{$^3$Department of Physics, Babes-Bolyai University, Kogalniceanu Street,
Cluj-Napoca 400084, Romania}
\affiliation{$^4$Department of Mathematics, University College London, Gower Street,
London, WC1E 6BT, United Kingdom}

\date{\today }

\begin{abstract}
The discovery of a large number of supermassive black holes (SMBH) at redshifts $z > 6$, when the Universe was only 900 million years old, raises the question of how such massive compact objects could form in a cosmologically short time interval. Each of the standard scenarios proposed, involving rapid accretion of seed black holes or black hole mergers, faces severe theoretical difficulties in explaining the short-time formation of supermassive objects. In this work we propose an alternative scenario for the formation of SMBH in the early Universe, in which energy transfer from superconducting cosmic strings piercing small seed black holes is the main physical process leading to rapid mass increase. As a toy model, the accretion rate of a seed black hole pierced by two antipodal strings carrying constant current is considered. Using an effective action approach, which phenomenologically incorporates a large class of superconducting string models, we estimate the minimum current required to form SMBH with masses of order $M = 2 \times 10^{9}M_{\odot}$ by $z = 7.085$. This corresponds to the mass of the central black hole powering the quasar ULAS J112001.48+064124.3 and is taken as a test case scenario for early-epoch SMBH formation. For GUT scale strings, the required fractional increase in the string energy density, due to the presence of the current, is of order $10^{-7}$, so that their existence remains consistent with current observational bounds on the string tension. In addition, we consider an ``exotic" scenario, in which an SMBH is generated when a small seed black hole is pierced by a higher-dimensional $F-$string, predicted by string theory. We find that both topological defect strings and fundamental strings are able to carry currents large enough to generate early-epoch SMBH via our proposed mechanism.
\end{abstract}

\maketitle


\section{Introduction} \label{Sect1}

The recent discovery of the ultra-luminous quasar SDSS J010013.02+280225.8 at redshift $z = 6.30$, with a central black hole mass of order $M=1.2\times 10^{10}M_{\odot}$ \cite{Q1}, where $M_{\odot}=2\times 10^{33}$ g is the Solar mass, has again raised the important theoretical question of how supermassive black holes (SMBH) can form and grow in the early Universe during astrophysically short time intervals. To date, around forty SMBH have been detected at  $z>6$ \cite{Q1}, with some at redshifts greater than seven. In particular, the quasar ULAS J112001.48+064124.3, located at $z=7.085$, has a central black hole of mass $M=2 \times 10^9 M_{\odot}$ \cite{Q2}. According to standard formation paradigms, involving accretion from the interstellar medium (ISM) and/or mergers, it is not possible for such a massive compact object to have existed when the Universe was only 0.77 billion years old.

Usually, it is assumed that SMBH grow from primordial seed black holes. In one approach, it is assumed that Pop III stars can produce black hole seeds with masses in the range $10-10^3M_{\odot}$ \cite{Q3}. This process may have started around $z\approx 20$. A second scenario involves the collapse of core regions in protogalactic disks at $z<12$, which could produce black hole seeds with masses of order $10^5-10^6M_{\odot}$ \cite{Q4}. Recently, an alternative mechanism, in which ultra-massive seeds are formed from the collapse of ``patches" in the early Universe, that experience greater than average Hubble expansion during the inflationary era, has also been proposed \cite{Nakama:2016kfq}.

Regardless of how the initial seed black holes may have formed, some type of mass accretion must play an important role in black hole growth. For accretion from the ISM, the mass growth rate can be modeled as \cite{Shap}
\be\label{1}
\frac{dM}{dt}=\frac{\epsilon _L\left(1-\epsilon _M\right)}{\epsilon _M}\frac{M}{\tau}.
\ee
The parameters $\epsilon _L$, $\epsilon _M$ and $\tau$ are given by $\epsilon _L=L/L_E$, $\epsilon _M=L/\dot{M}_0c^2$ and $\tau =Mc^2/L_E$, where $\dot{M}_0$ denotes the rate of rest-mass accretion, $L$ is the luminosity, and $L_E=4\pi \mu _em_pc/\sigma _T$ is the Eddington luminosity. The latter is defined in terms of  the proton mass $m_p$, the Thompson cross section $\sigma _T$, and the mean molecular weight per electron $\mu _e$, which is related to the primordial helium abundance $Y$ according to $\mu _e=1/(1-Y/2)$, with $Y=0.25$.

Other possible explanations for the ``Impossible Early Galaxy Problem" include failed template fitting or errors in cosmological  redshift determination, early star formation, or the presence of new clustering physics \cite{Imp}. Unfortunately, the precise details of the merging process(es), or of the clustering, are not known, so that alternative possibilities that could explain the observational data can not be rejected {\it a priori}. In the present paper, we outline a scenario in which cosmic strings $-$ line-like defects predicted to form during phase transitions in the early Universe \cite{VS_book,cosmic_strings} $-$ are able to play an important role in the dynamical evolution of SMBH.

Theoretically, it has been suggested that small numbers of strings may have survived to the present day, though observational bounds limit the long string number density to $\sim \mathcal{O}(1)$ per horizon volume at the current epoch \cite{VS_book}. Nonetheless, it is possible that their existence in the early Universe may help to explain some intriguing astrophysical problems. For example, strings formed at a symmetry breaking phase transition associated with the with grand unification (GUT) scale, of order $10^{16}$ GeV, could have induced density fluctuations that acted as seeds for the formation of galaxies, despite being ruled out as the primary source of primordial cosmological perturbations \cite{Pe}. Since the mass per unit length of the strings is determined by the symmetry breaking energy scale at which they are formed, the gravitational effects of both long strings and  loops are nontrivial, and strings may act as gravitational lenses \cite{Gott:1984ef}, source subdominant temperature fluctuations in the CMB \cite{CS_CMB}, and induce $B$-mode polarization \cite{CS_B-mode}.

Recently, it has also been suggested that the massive compact seeds required at very high redshifts, in order to generate SMBH using standard accretion scenarios, may have been provided by cosmic string loops \cite{Bram}. If string loops were present in the early Universe then, by gravitational accretion, they could have led to the formation of ultra massive objects. However, it is the purpose of this work to propose an alternative model for the formation of SMBH at early epochs. Specifically, we show that a small seed black hole pierced by superconducting strings can experience very rapid growth due to energy transfer from the string current.

Our general approach is based on an important generalization of the original (vacuum) cosmic string theory given by Witten \cite{Witten85}, who first proposed that
strings may carry electric currents, thus behaving like superconducting wires. The charge carriers may either be bosons, in which case a charged Higgs field with a nonzero vacuum expectation value in the core of the string is required, or fermions, which are trapped as zero modes along the string. In \cite{Witten85}, superconducting strings were described by a toy $U(1) \times \tilde{U}(1)$ model, though many additional superconducting string models have since been proposed in the literature. (See \cite{Sabancilar:2013wta} for a recent review of superconducting string phenomenology). By modifying the estimate of the string current, given in \cite{Witten85}, to include redshift dependence, we show that these ``chiral" strings are able to provide the dominant contribution to early-epoch SMBH growth when a small seed black hole intersects the string network. 

Strictly, we extrapolate the results obtained in \cite{Witten85} to estimate the current induced, in either an oscillating string loop, or in a section of long string moving at relativistic velocity, in the presence of a primordial magnetic field. Thus, this model also includes the important case of current-induction due to long string motion in the context of the velocity-dependent one-scale (VOS) model \cite{VOS}, which provides the best analytic description of the evolution of the large-scale properties of string networks. In both cases, we find that the induced current is sufficiently high for the proposed mechanism to form the dominant contribution to SMBH accretion, even when energy loss due to gravitational and/or electromagnetic radiation from the strings is taken into account. We then adopt a more general phenomenological approach, based on the effective action for a superconducting string derived in \cite{CS_EFF.ACT}, which is valid for a large class of individual string models.

Though this macro-description necessarily neglects the effects of small-scale structure, including sharp kinks and/or propagating ``wiggles", which may also have important consequences for the the evolution of the string network \cite{small-scale-structure}, it is the large-scale properties that are of greatest relevance to our model. Since the equation of state for wiggly strings is, in any case, similar to that for superconducting strings \cite{small-scale-structure}, our approach remains valid. In addition, the presence of very large, high momentum density, currents on the string may be expected to ``smooth out" small-scale kinks. An extreme example of this is represented by the cosmic ``vorton" scenario \cite{vortons}, which is discussed separately in Sec.~\ref{Sect5}.

An interesting astrophysical model, incorporating both strings and black holes, was proposed in \cite{Ary86}. In this, a complex gravitational system consisting of a Schwarzschild black hole with a straight string passing through it was analyzed. A black hole with strings emanating from it could have been formed during a phase transition in the post-inflationary era. One possible type of string, which may be considered as a candidate for the black hole$-$string system, is a flux tube of confined gauge fields. Above the phase transition temperature, a black hole containing a nonzero electric charge will have a spherically symmetric Coulomb-type field \cite{Ary86}. If, due to cosmological expansion, the system cools to below the critical temperature, the Coulomb field will become confined and strings emanating from the central black hole can be formed.

Gauss's theorem requires that the total energy flux carried by the strings must be equal to the net energy flux across the black hole's horizon before the transition. In general relativity, classical  fields such as the electromagnetic or Yang-Mills fields, which satisfy Gauss's law, do not violate the black hole ``no hair" theorems. (See \cite{Bekenstein:1996pn} for a review.) Nonetheless, a black hole formed in the early Universe may have ``hairs'' in the form of strings, as shown in \cite{Ary86}. The extension of the Schwarzschild metric for a black hole pierced by a vacuum cosmic
string is
\begin{eqnarray}  \label{2}
ds^2&=&-\left(1-\frac{2G}{c^2}\frac{\mathcal{M}}{r}\right)c^2dt^2+\left(1-\frac{2G}{c^2}\frac{\mathcal{M}}{r}\right)^{-1}dr^2
\nonumber \\
&+& r^2\left(d\theta ^2+b^2\sin ^2\theta d\phi ^2\right),
\end{eqnarray}
where $\mathcal{M} = bM$ is the effective mass and $M$ is the physical mass. Here, $b^2 = 1-\delta$, where $\delta = 8\pi G \mu/c^2$ is the deficit angle induced by the string and $\mu$ is the string mass per unit length. A cosmic string possesses a positive energy density so that $\mu >0$ and $b<1$. This initial model of a black hole pierced by strings has been extended to include the cases of charged and rotating black holes, and the properties of these composite systems have been intensively investigated \cite{refp}. However, to date, no model of mass accretion due to superconducing strings piercing the horizon of a black hole has been constructed. We here present one and consider its astrophysical and cosmological implications.

Ironically, one of the greatest difficulties in developing a theory of black hole mass accretion from superconducting strings is choosing a specific string model. Over the past forty years, cosmic strings have been extensively studied by theorists and a huge number of different models now exist. Broadly speaking, these can be categorized into two classes: topological detect strings, predicted by quantum field theories in which axial symmetry is spontaneously broken \cite{VS_book}, and ``$(p,q)-$strings'', bound states of $p$ macroscopic fundamental strings (or $F-$strings) and $q$ $D1-$branes (or $D-$strings), predicted by string theory \cite{pq_strings}. For $p>1$, $q>0$, these strings may become superconducting, since it has been shown that forming a $(p,q)$ bound state is equivalent to dissolving $p-1$ units of electric flux and $q$ units of magnetic flux on the world-sheet of a single $F-$string \cite{pq_strings}.

In models of the early Universe motivated by superstring theory \cite{superstrings}, networks of current carrying cosmic strings are predicted to form at the end of brane inflation \cite{CS_brane_infl.}. Nonetheless, even if networks of $F-$strings $-$ described by the Nambu-Goto action \cite{Nambu-Goto} in which the strings carry no additional world-sheet fluxes $-$ are created instead, these may still carry charge density and current, from a four-dimensional perspective, due to their motion in the compact internal space \cite{dim_red}. As the existence of extra dimensions is unavoidable in string theory, this possibility cannot be discounted \emph{a priori} if we are to take string theory seriously as candidate Theory of Everything (TOE). Further complications also arise from the fact that, within the vast array of current-carrying string models, currents may be time-like or null, and strings may carry nonzero net charge or be charge neutral (see, for example \cite{CS_EFF.ACT}).

For the sake of concreteness, we choose to model superconducting strings with null current and nonzero net charge, though it is important to note that  qualitatively similar results should hold for black holes pierced by strings carrying time-like currents, including neutral strings. Within these two limitations, we would like our model to be as general as possible. We therefore focus on general features of string phenomenology, which are common to virtually all superconducting string species. We aim to develop a model that is \emph{phenomenologically robust} and applicable to a large class of specific string models, irrespective of the precise underlying theory of string formation/evolution.

To this end, we consider the nature of superconducting strings from three different viewpoints. First, we consider the general relation between the current and the mass of the charge carriers, using this to directly obtain an estimate for the mass accretion rate of the seed black hole in terms of the string current, $J$. (However, this estimate may be considered simplistic, in the sense that it accounts only for the rest mass of the charge carriers and not, explicitly, for the additional mass transfer due to their kinetic energy.) As an example, this is evaluated for chiral strings, using an estimate of the bosonic current originally derived in \cite{Witten85}, and it is found that, in this test-case, the current is sufficient to provide the dominant mass accretion mechanism for the generation of early-epoch SMBH.

Second, we use the effective action developed by Copeland, Turok and Hindmarsh (CTH) \cite{CS_EFF.ACT}, which is valid for superconducting strings with either bosonic or fermionic currents, to determine the nonzero components of the string energy-momentum tensor. Since these determine the flow of energy along the string (which, in our toy antipodal string model, is converted into black hole rest mass when it passes the horizon), we obtain improved estimates for the mass accretion rate, for different values of $\mu$ and the dimensionless current $j \propto J$.

Third, we consider an ``exotic" scenario, motivated by string theory, in which the black hole is pierced by an $F-$string, which exists in a higher-dimensional space-time. For a higher-dimensional black hole with Schwarzschild radius $r_S \gg R$, where $R$ is the radius of the compact space, the metric is effectively four-dimensional \cite{Horowitz:2012nnc} so that the main physical effect of the extra dimensions on the string$-$black hole system arises from the motion of the string in the internal directions. Under dimensional reduction, this gives rise to an effective world-sheet flux from a four-dimensional perspective, which may be interpreted as the flow of electric charge \cite{dim_red}. We explicitly show that, from a four-dimensional perspective, such an exotic model is directly equivalent to the effective action approach developed by CTH \cite{CS_EFF.ACT}, and we determine the field-theoretic model parameters in terms of the string theory model parameters, including $R$. Since, by the results obtained in \cite{dim_red}, current-carrying strings in $d$ spatial dimensions can be viewed as ordinary (non-superconducting) strings in $d+1$ spatial dimensions, the case of superconducting $(p,q)-$strings is automatically included in this approach. Hence, these may also be modelled by the effective action given by CTH \cite{CS_EFF.ACT}, as long as an appropriate expression for the effective ``intrinsic" tension of the string is used (see, for example, \cite{pq_strings} plus \cite{Lake:2010qsa} and references therein).

In the higher-dimensional model, the winding radius $R$ plays the role of the effective string width, from a $(3+1)$-dimensional perspective, and is equivalent to the core radius $r_c$ in field-theoretic models. As the existence of a finite string width has little physical impact on the dynamics of the string--black hole system, at least for $r_s \gg r_c \sim R$, we are justified in using the CTH action in this regime. This is a simple modification of the Nambu-Goto action, which is extended to include additional world-sheet fluxes, and is equivalent to taking the ``wire approximation" \cite{Anderson_book} for superconducting strings. The physical effects of the finite string core are therefore neglected, although the quantity $r_c$ (or equivalently $R$) appears in the expression for the string tension/mass per unit length via $\mu \sim 1/r_c^2$. Wherever there is a need to account explicitly for a finite core radius, such as in Sec.~\ref{Sect8}, where astrophysical realizations of our model are considered, this is explicitly stated.

This paper is organized as follows. In Sec.~\ref{Sect2}, we use order of magnitude estimates of the current and the mass of the charge carriers in a superconducting string to estimate the flow of energy. We introduce the phenomenological superconducting string model and the mass transfer into a black hole pierced by two long, ``straight" antipodal strings, with opposite charge density and current, is estimated. This is intended as a toy model for the creation of an uncharged and stationary SMBH, and we note that the curvature radius of the strings need only be ``long" compared to the radius of the black hole horizon. In reality, the seed black hole may be pierced by large loops, and the estimates obtained remain valid, so long as the loops do not expire before $z \approx 7$. To ensure this is indeed the case, the energy loss due to both gravitational and electromagnetic radiation from a string loop is also estimated and we find that the mechanism remains viable for realistic (GUT scale) strings. Hence, ÒstraightÓ, in this context, refers to a string -- possibly connected to a network, or possibly a loop -- whose curvature radius is large compared to the radius of the black hole it pierces, though this in no way precludes the existence of small-scale structure, which may play an important role in string evolution \cite{small-scale-structure}.
In Sec.~\ref{Sect3}, we refine our analysis by introducing the effective action for a superconducting string, which is valid for both bosonic and fermionic currents, and which allows us to calculate the accretion rate directly in terms of the string constants of motion, i.e., the conserved current and its associated energy/momentum. The mass transfer to a black hole pierced by two antipodal superconducting strings, described by the CTH action, and with opposite charge and current flow, is considered in Sec.~\ref{Sect4}. In Sec.~\ref{Sect5}, we consider an ``exotic" scenario, motivated by string theory, in which higher-dimensional $F-$strings with windings in the compact internal space replace the field-theoretic strings considered in Sec.~\ref{Sect3}. In Sec.~\ref{Sect6}, we comment briefly on physical mechanisms by which a composite system of black holes and superconducting strings may be formed in the early Universe. In Sec.~\ref{Sect7} we explore the astrophysical and cosmological implications of our previous models and several specific issues that must be addressed, if the proposed accretion mechanism is to occur in a realistic astrophysical scenario, are considered in Sec.~\ref{Sect8}. A summary of our main conclusions and a brief discussion of prospects for future work are given in Sec.~\ref{Sect9}.

\section{Simple estimates of the string current and mass accretion rate of a seed black hole} \label{Sect2}

For a string carrying constant current $J$, we may write
\be \label{J-1}
J = \frac{Q}{t} = \frac{n_q q}{t},
\ee
where $Q$ is the total charge passing a point on the string within time $t$, $n_q$ is the number of charge carriers, and $q$ is the charge of an individual carrier. If the charge carriers have mass $m_q$, we may estimate the total mass transfer to a black hole pierced by two antipodal strings as
\be \label{M-1}
M(t) = 2m_qn_q = 2\frac{m_qJ}{q}t.
\ee

In order to evaluate the expression on the right hand-side of Eq. (\ref{M-1}), we need to know the fundamental charge and mass of the carriers and to estimate the current. An estimate of the current carried by a superconducting string with bosonic charge carriers may be obtained by considering the movement of the string in an external magnetic field. For strings moving a distance $l$ in an external magnetic field $B(z)$, the corresponding current can be estimated as
\bea \label{J-B}
J &=& \frac{2\pi}{\ln \left(c/(r_c\omega_s) \right)}B(z) c l = 1.9\times 10^{16}
\nonumber\\
&\times& \left(\frac{B(z)}{10^{-6}\;{\rm G}}\right)\left(\frac{l}{10^4\;{\rm pc}}\right)\left(\frac{100}{\ln \left[c/(r_c\omega_s)\right]}\right)\;{\rm A},
\eea
where $\omega_s$ is the characteristic frequency of the string motion and $r_c$ is the width of the string core \cite{Witten85}. In the above equation, $z$ is the cosmological redshift and we may assume that, beyond the damping scale, the magnetic field scales as $B(z) \sim B_0/a^2(z) \sim B_0(1+z)^2$, where $B_0$ is the present day field strength and $a(z)$ is the scale factor of the Universe. This scaling is required by flux conservation \cite{magn}.

The characteristic values $l \sim 10^4$ parsecs and $B_0 \sim 10^{-6}$ G correspond, roughly, to the radius of the luminous portion of an average galaxy, and to the strength of the coherent component of the present day ($z=0$) galactic magnetic field, respectively \cite{Witten85}. Generally, we may also assume that the strength of the present day galactic magnetic field is of the same order of magnitude as the inter-galactic field strength, which may be primordial in origin \cite{magn}.
 
Using (\ref{M-1}), we obtain
\bea \label{M-B}
M(t) &=& 4.228 \times 10^{27}\times (1+z)^2
\nonumber\\
&\times& \left(\frac{m_q}{1.782 \times 10^{-8}\;{\rm g}}\right)  \left(\frac{1.603 \times 10^{-19} \ {\rm C}}{q} \right)^{-1}
\nonumber\\
&\times&  \left(\frac{B_0}{10^{-6}\;{\rm G}}\right)\left(\frac{l}{10^4\;{\rm pc}}\right)\left(\frac{100}{\ln \left[c/(r_c\omega_s)\right]}\right) t \ {\rm g}.
\eea
as an estimate for the total mass transferred, in time $t$, from the string to the black hole, where we have used $e = 1.603 \times 10^{-19} \ {\rm C}$ and $m_{\rm GUT} \approx 10^{16} \ {\rm GeV/c^2} = 1.782 \times 10^{-8} {\rm g}$ as reference scales for the charge and mass, respectively, and where $\ln \left[c/(r_c\omega_s)\right] \sim 100$ corresponds to characteristic values of $\omega_sr_c$ at the present epoch \cite{Witten85}. 

For $q=e$, $m_q = m_{\rm GUT}$, and characteristic values of $l$, $B_0$ and $r_c\omega_s$, this gives $J \approx 1.216 \times 10^{18}$ A and $M \approx 6.719 \times 10^{45} {\rm g} = 3.358 \times 10^{12} M_{\odot}$ for $t = 2.43 \times 10^{16}$ s ($0.77$ billion years). In a cosmological context, we also note that this value of $l$ corresponds to the size of the cosmological horizon, $l_{\rm H}(z)$, at $t \sim 10^{12}$ s ($z \gtrsim 650$). In other words, if the string moves a distance $l \sim 10^{-4}l_{\rm H}(z=7.35)$, corresponding to an average velocity $\langle v\rangle \sim 10^{-4}c$ between the epoch of string formation (assumed to be $t_s \approx 0$) and the epoch corresponding to the observation of our test-case SMBH, this will induce a sufficiently large current to generate an SMBH from a seed black hole. In principle, an {\it average} current of this magnitude may be induced in a section of long string, connected to the string network but moving at relativistic velocities $\langle v\rangle \gtrsim 10^{-4}c$, or in an oscillating string loop.

In the VOS model of string network evolution \cite{VOS}, which accurately matches the results of numerical simulations \cite{VOS_simul.}, the evolution of the string correlation length is determined by Hubble friction, the damping scale (due to collisions of the string with background plasma), the loop chopping efficiency parameter $\tilde{c}$, and the average long-string velocity $v_{\infty}$. Frictional damping dominates over a time scale $t_{*} \sim (G\mu/c^2)^{-1}t_s$, where $t_s$ is the time of string formation, and is thus negligible for GUT-scale strings, which are able to reach relativistic velocities at very early times \cite{VOS}. Hence, even using a conservative estimate, based on the current generated by long string motion and/or loop oscillations with characteristic velocity $\langle v \rangle \sim 10^{-4}c$, in the presence of the primordial magnetic field present {\it at} $z=7.085$, the mass accretion from superconducting GUT strings piercing seed black holes is at least {\it three} orders of magnitude larger than that required to form the SMBH powering ULAS J112001.48+064124.3. In reality, the primordial magnetic field should have been even stronger for $z \gtrsim 7$.

Of course, this estimate of the mass accretion may be significantly affected by uncertainties in the values of the present day galactic magnetic fields, as well as by the complexities of the cosmological evolution of $B(z)$. Moreover, it should be noted that this scenario {\it assumes} the following: (a) that the seed black hole intersects the string network at, or soon after, the epoch of string formation, (b) that string formation occurs in the very early Universe at $t \approx 0$, (c) that the black hole remains ``plugged in'' to the string network over a time scale of (at least) $0.77$ billion years, and (d) that the seed black holes have negligible initial mass. The second assumption is reasonable for GUT scale strings, since $t_{\rm GUT} \approx 10^{-7}$ s, and the fourth represents the most conservative scenario for the seed masses of early epoch SMBH. By contrast, the validity of both the first and third assumptions is debatable. However, we also note that, even if this mechanism is only 0.1$\%$ efficient, it is still capable of creating SMBH in the required mass range at redshifts $z \approx 7$ and, if it is only 0.01$\%$ efficient, may still form the dominant accretion mechanism for SMBH growth.

Furthermore, the estimates above may be compared with those obtained for the maximum threshold current, for either bosonic or fermionic strings, above which the strings become unstable and decay \cite{Witten85},
\be \label{Jmax}
J_{\rm max} = q\frac{m_q c^2}{2\pi \hbar}.
\ee
For charge carriers with $q=e$ and masses comparable to the electron mass $m_e = 9.109 \times 10^{-28}$ g the maximum current carried by a single zero mode is of the order of $20$ A, whereas for GUT scale particles it is $4 \times 10^{20}$ A \cite{Witten85}. Combining this expression with Eq. (\ref{M-1}), we have
\be \label{M-Jmax}
M(t) = \frac{m_q^2c^2}{\pi\hbar}t,
\ee
giving $M = 2.093 \times 10^{48}$ g $= 1.047 \times 10^{15} M_{\odot}$ at $z = 7.085$ for our previous choice of parameters. Thus, parameterizing the time-averaged current in terms of the maximum threshold value, such that
\be \label{<J>}
\langle J \rangle = \epsilon J_{\rm max}
\ee
with $0 \leq \epsilon \leq 1$, we see that, for GUT scale particles, we require only $\epsilon_{\rm GUT} \sim \mathcal{O}(10^{-6})$. Alternatively, for currents close to the threshold value, the accretion mechanism outlined here need only be of order $10^{-4}\%$ efficient in order to generate the observed SMBH at $z \approx 7$. Such inefficiency could be caused by a number of factors, including late-time interaction of the seed black hole with the cosmic string network and/or premature decoupling of the string$-$black hole system. For example, if a seed black hole interacts with a GUT scale string carrying the threshold current, it need only accrete mass for approximately 1000 years. It is therefore important to estimate the average lifetime of the superconducting strings which may intersect the seed black holes.

Before concluding this subsection, we note that, in our toy model, in which antipodal strings carry oppositely charged currents, with equal magnitude and opposite direction, the net magnetic flux through the black hole horizon is zero. Hence, the black hole carries no magnetic monopole charge. The conventional current ``through" the black hole is also constant, so that no net electric charge is accumulated.

\subsection{String loops -- the effects of radiative emission} \label{Sect2.1}

The lifetime of a (non-self-intersecting) string loop is determined by the rate at which it radiates energy, either in the form of gravitational or electromagnetic waves. For macroscopic loops, gravitational radiation is the main energy-loss mechanism \cite{VS_book}. The gravitational radiation power $\dot{E_{G}}$ for a loop of length $l$ can be roughly estimated using the quadrupole formula for gravitational waves, and is given by
\be
\dot{E}_G=\Gamma _{G}G\mu ^2c,
\ee 
where $\mu $ is the mass per unit length of the string. The numerical coefficient $\Gamma _G$ is independent of the loop size, but may depend on its shape and trajectory \cite{VS_book}. The gravitational lifetime of the loop $\tau _G$ is then
\be
\tau _G\sim\frac{l \mu c^2}{\dot{E}_G}\sim \frac{lc}{\Gamma _GG\mu}. 
\ee
Assuming that the gravitational properties of the string are characterized by the dimensionless quantity $G\mu/c^2 \sim 10^{-6}$, which corresponds to GUT scale strings, and by taking $\Gamma _G\sim 100$ \cite{VS_book}, and assuming a scaling solution, $l(t) = \alpha_s c t$, where $\alpha_s \lesssim 10^{-3}$ \cite{VS_book}, this rough estimate of the gravitational life-time of the string gives $\tau_G \sim 10t$, where $t$ is the epoch of loop formation.

Hence, we see that, if the the seed black hole becomes plugged into a section of the GUT string network at very early times and that section remains, effectively, ``long" (i.e. does not chop off to form a loop, which then decays via gravitational radiation), for a time $\Delta t \gtrsim 10^{16}$ s, it is easy to transfer sufficient mass-energy to the black hole. Specifically, this happens if the current in the network is greater than, or comparable to, $0.1\%$ of the estimate given by Eq. (\ref{J-B}) or, equivalently, $10^{-4}\%$ of the threshold current. On the other hand, if the local section of the long-string network to which the seed black hole is attached chops off to form a loop at $t \lesssim 10^{12}$ s, for GUT scale strings carrying the characteristic current given by Eq. (\ref{J-B}), or at  $t \lesssim 10^{9}$ s, for strings carrying the GUT scale threshold current, it will decay due to gravitational wave emission {\it before} it is able transfer enough mass to form the dominant contribution to SMBH accretion. 

Roughly speaking, when $t$ is reduced by one order of magnitude, the contribution to SMBH accretion is also reduced by one order of magnitude. Hence, for loop creation in the scaling regime, the ultimate viability of our model may depend sensitively on the detailed time evolution of the loop production parameter, commonly denoted $\tilde{c}$ \cite{VS_book}, though, since only $\sim 10$-$100$ string-fueled accretion events must occur, in order for this mechanism to account for the {\it current} observational bounds on the number of SMBH at $z \gtrsim 7$, it remains a viable alternative to existing models at the present time. 

However, we note that, while both analytical models and numerical simulations suggest that the inter-string distance scales from early times, the most recent (and accurate) simulations of Nambu-Goto string networks suggest a much slower approach to scaling for loops \cite{BlancoPillado:2011dq}. Though the reasons for this remain unclear, the most likely explanation is the fragmentation of the initial loop due to the presence of large kinks on the string, as suggested long ago \cite{loop-frag}. However, again, since only $\sim 10$-$100$ seed black hole--string systems are required in order to satisfy current observational constraints, we may use the scaling solution as an upper bound on the loop size, and consider only string--black hole systems of this, or comparable, scales. Though, in principle, many seed black holes may be connected to much smaller sections of string (loops), these will be unable to transfer significant mass to the seed to form SMBH and may be discounted in our model. Since the question of the (non-)scaling of loops at all redshifts \cite{Lorenz:2010sm} remains a controversial point in the existing literature \cite{loop-non-scale}, our assumption of a scaling must be considered as a (somewhat crude) first approximation and, most importantly, as giving rise only to an upper bound on the loops size, and to a lower bound on the number of sufficiently large string--black hole systems.

For superconducting strings, energy loss due to electromagnetic radiation must also be taken into account. A current-carrying loop, oscillating in vacuum, may be a powerful source of electromagnetic radiation. For a loop without kinks or cusps, the power $\dot{E}_{em}$ emitted through electromagnetic processes is \cite{VS_book}
\be
\dot{E}_{em}=\frac{\Gamma _{em}}{c}J^2.
\ee
By again assuming a current parametrization of the form (\ref{<J>}), we obtain
\be
\dot{E}_{em}=\epsilon ^2\Gamma _{em}\frac{c^3}{(2\pi\hbar)^2}m_q^2q^2,
\ee
so that the mass lost through electromagnetic radiation is 
\be \label{emmass}
\frac{dM_{em}}{dt}=\epsilon ^2\Gamma _{em}\frac{c}{(2\pi\hbar)^2}m_q^2q^2.
\ee
For $\Gamma _{em}=100$ and assuming, for the sake of consistency, that the charge carriers have fundamental charge $e$ and mass $m_{\rm GUT} \sim 10^{-8}$ g, Eq.~(\ref{emmass}) gives $dM_{em}/dt \sim 10^{-5} \epsilon^2$ g/s ($dM_{em}/dt \sim 10^{-28} \epsilon^2 \rm M_{\odot}/s$), which for a time span of 0.77 billion years gives a total mass loss of $M_{\rm loss} \sim 10^{-12} \epsilon^2 M_{\odot}$.

As we have already mentioned, an efficiency of the order of $\epsilon_{\rm GUT} \sim 10^{-6}$ is sufficient to make the present process effective for GUT scale strings, neglecting energy loss through radiative effects. However, even in the extreme limit $\epsilon_{\rm GUT} \rightarrow 1$, the mass loss through electromagnetic radiation is extremely low, and does not significantly affect the black hole mass increase. Significantly larger values of $\Gamma _{em}$ do not affect the results of this analysis. Therefore, we see that the emission of electromagnetic radiation from superconducting GUT strings plays a negligibly small role in the determining string loop lifetimes. As a result, it has no significant impact of the formation of SMBHs in the early Universe via the mechanism proposed here.

Note that similar conclusions, regarding loop lifetimes, and the effects of both gravitational and electromagnetic radiation, also hold for more general superconducting string models, including those described by the effective action introduced in \cite{CS_EFF.ACT}, which is used throughout the rest of this paper. As such, the results presented herein remain valid, subject to the requirement that, if the seed black hole is pierced by loops, these must be sufficiently ``large". In other words, they must be large enough to be indistinguishable, from the perspective of the accreting seed black hole, from genuine ``long" string sections, throughout the time required for SMBH formation. However, for the sake of simplicity in the following analysis, we {\it assume} that the ``long" string--black hole system survives for at least a time period $\Delta t \sim 10^{16}$ s. Nonetheless, even if this assumption is relaxed, the preceding arguments demonstrate that the SMBH growth model presented here remains viable, for sufficiently high string currents, even on much shorter time scales. 

A further complication arises regarding the (non-)applicability of the loop scaling solution to field-theoretic strings. In recent work by Hindmarsh, Stuckey and Bevis \cite{Hindmarsh:2008dw}, it is claimed that the discrepancies between simulations of Nambu-Goto and Abelian-Higgs string networks may be resolved if (a) horizon-sized loops fragment, as suggested by Scherrer and Press (among others) \cite{loop-frag} and (b) energy is radiated from loops at a constant rate. Since the loop size distribution obtained by Scherrer and Press is logonormal, this implies that a significant number of sufficiently large loops, whose radius is proportional to $t$, may still be formed. However, the prediction of linear energy loss from loops contradicts the standard analysis of radiative emission presented in Sec.~\ref{Sect2.1}, and has potentially more serious consequences for our model. Despite this, we note that, even in the case of linear energy loss, our mechanism remains viable so long as the associated power is smaller than the power associated with the string current. If both energy loss from the string to the environment and to the black hole are linear, even such a drastic modification of the usual scenario does not invalidate our hypothesis, but merely reduces its efficiency.

Finally, we note that the estimate for the mass transfer (\ref{M-1}) may be considered simplistic, in the sense that it accounts only for the total \emph{rest mass} of the charge carriers crossing the black hole horizon and not their kinetic energy. For our toy model of accretion from strings, in which the net momentum transfer to the black hole is zero, the kinetic energy of the current is also converted into black hole rest mass when it crosses the horizon and, in principle, this may provide a nontrivial contribution to the total accretion rate. To determine whether this is indeed the case, we must develop a completely relativistic treatment of the current and obtain an exact expression for the momentum $P$, corresponding to a given $J$. This is obtained in the next Section, using the effective action approach developed by CTH \cite{CS_EFF.ACT}.

\section{Estimate of the accretion rate from the effective action of a superconducting string} \label{Sect3}

In this Section, we review the basic electromagnetic properties of long, straight, superconducting strings. The approach is phenomenological and the results are valid for a large class of superconducting string models, carrying either bosonic or fermionic currents. We begin by reviewing the derivation of the effective action for a vacuum string (with zero current) before generalizing this to the superconducting case. Using these general results, we give order of magnitude estimates for the accretion rate of a seed black hole in our toy model, in which the black hole is pierced by two antipodal strings carrying currents with opposite charge, flowing in opposite directions. Assuming that the string$-$black hole system forms in the very early Universe (i.e. close to $t=0$ on a cosmological time scale), we estimate the current required to form SMBH with masses of order $M=2 \times 10^9 M_{\odot}$ within a time interval of $0.77$ billion years. This corresponds to the observed quasar ULAS J112001.48+064124.3, which we take as a test case.

\subsection{Microphysical ansatz and effective action for a vacuum string} \label{Sect3.1}

The simplest field-theoretic description of a cosmic string arises in the Abelian-Higgs model, given by the action
\begin{eqnarray} \label{AH_act}
S = \int \sqrt{-g}d^4x\left\{D_{\mu}\phi D^{\mu*}\phi^{*} - \frac{1}{4}F_{\mu\nu}F^{\mu\nu} - V(|\phi|)\right\}
\end{eqnarray}
where $\phi$ is a complex scalar field with potential
\begin{eqnarray} \label{AH_poy}
V(|\phi|) = \frac{\lambda}{4}(|\phi|^2 - \eta^2)^2.
\end{eqnarray}
(For convenience, we set $c = \hbar = G = 1$ throughout this section.) The gauge covariant derivate is defined as $D_{\mu} = \partial_{\mu} - ieA_{\mu}$ and the electromagnetic field tensor is defined, as usual, as $F_{\mu\nu} = \partial_{\mu}A_{\nu} -  \partial_{\nu}A_{\mu}$. The action (\ref{AH_act}) is invariant under local $U(1)$ gauge transformations giving rise to a conserved current $j^{\mu} = -ie[\phi^{*}D^{\mu}\phi - \phi D^{\mu*}\phi^{*}]$, in addition to the conserved energy-momentum tensor
\begin{eqnarray} \label{AH_EMT}
T^{\mu\nu} = D^{\mu}\phi D^{\nu*}\phi^{*} +  D^{\mu*}\phi^{*}D^{\nu}\phi - F^{\mu}{}_{\alpha}F^{\nu\alpha} - g^{\mu\nu}\mathcal{L}.
\end{eqnarray}

Physically, the parameters $\lambda$ and $e$ denote the scalar and vector field couplings, respectively, while $\eta$ represents the energy scale at which the local $U(1)$ symmetry of the vacuum is spontaneously broken. This allows for the formation of vortices/strings described by the following ansatz, first proposed by Nielsen and Olesen \cite{Nielsen:1973cs},
\begin{eqnarray} \label{NO_ansatz}
\phi(r,\theta) = \eta f(r)e^{in\theta}, \ \ \ A_{\theta}(r) = \frac{n}{e}a_{\theta}(r)
\end{eqnarray}
with $A_0 = A_z = 0$, and $A_r = 0$, where we have used cylindrical polar coordinates $\left\{t,r,\theta,z\right\}$ and assumed a local Minkowski metric within the string core.
The topological winding number $n \in \mathbb{Z}$ is given by
\begin{eqnarray} \label{NO_n}
n = \frac{1}{2\pi}\int_{0}^{2\pi} \frac{\partial \Theta}{\partial \theta}d\theta
\end{eqnarray}
 where $\Theta$ denotes the argument of the scalar field $\phi$ and $f(r)$, $a_{\theta}(r)$ are dimensionless functions obeying the boundary conditions
 \begin{eqnarray} \label{NO_f_bc}
f(r) = \left \lbrace
\begin{array}{rl}
0,& \ r=0 \\
1,& \ r \rightarrow \infty,
\end{array}\right.
\
a_{\theta}(r) = \left \lbrace
\begin{array}{rl}
0,& \ r=0 \\
1,& \ r \rightarrow \infty.
\end{array}\right.
\end{eqnarray}
By substituting the ansatz (\ref{NO_ansatz}) into the covariant EOM, it may be shown that, to leading order
\begin{eqnarray} 
f(r) \approx \left \lbrace
\begin{array}{rl}
(r/r_{s})^{|n|},& \ r \lesssim r_{s} \\
1,& \ r \gtrsim r_{s},
\end{array}\right.
\
a_{\theta}(r) \approx \left \lbrace
\begin{array}{rl}
(r/r_{v})^{2},& \ r \lesssim r_{v} \\
1,& \ r \gtrsim r_{v},
\nonumber
\end{array}\right.
\end{eqnarray}
\begin{eqnarray} 
{}
\end{eqnarray}
where $r_s$ and $r_v$ denote the scalar and vector core radii, respectively. These are given by the inverse masses of the associated bosons \cite{VS_book}
\begin{eqnarray} \label{NO_core_radii}
r_s = m_s^{-1} \approx (\sqrt{\lambda}\eta)^{-1}, \ \
r_v = m_v^{-1} \approx (\sqrt{2}e\eta)^{-1}.
\end{eqnarray}
Substituting the approximate solutions (\ref{NO_f}) into (\ref{AH_EMT}), the only nonzero components of the energy-momentum tensor are $T^{00}$ and $T^{zz}$. It is straightforward to show that, at critical coupling, defined as
\begin{eqnarray} \label{crit_couple}
r_v^2/r_s^2 = \lambda/(2e^2) = 1 \iff  r_v = r_s = r_c,
\end{eqnarray}
where $r_c$ denotes the (equal) scalar and vector core radii, the energy and integrated pressure (i.e. tension) of a finite section of string of length $L$ are, respectively,
\begin{eqnarray} \label{NO_ET}
E = \mu L, \ \ \ T^{z} = -\mu L,
\end{eqnarray}
where $\mu \approx 2\pi\eta^2|n|$ is the string mass per unit length. Hence, we see that, for a vacuum string, the local string tension $\mathcal{T}^z$ and mass per unit length $\mu$ are related via $\mathcal{T}^z = -\mu$. It is also straightforward to show that $j^{\nu} = 0$ for all $\nu$, so that the string is uncharged and does not carry a current.

An effective action for the Nielsen-Olesen string can be constructed by switching to a set of dimensionless world-sheet coordinates $\zeta^{a}$, $a \in \left\{0,1\right\}$, where $\zeta^0 = \tau$ is time-like and $\zeta^1 = \sigma$ is space-like. These parameterize the two-dimensional sheet swept out by the line $\langle \phi \rangle = 0$, which represents the central axis of the string core. Erecting two normals to this sheet, denoted $n_{\alpha}{}^{\mu}$, we may describe a small region around the string core using the coordinates $Y^{\mu} = X^{\mu}(\tau,\sigma) + n_{\alpha}{}^{\mu}\rho^{\alpha}$, where $X^{\mu}$ are the embedding coordinates of the core central axis and $\rho^{\alpha}$ probe the core region $r \in [0,r_v]$ \cite{CS_EFF.ACT}. (Here we assume $r_s \leq r_v$, which corresponds to a Type II superconductor regime \cite{VS_book}.) If the curvature radius of the string is much greater than its thickness, we may approximate the integral in the action (\ref{AH_act}) as $d^4x\sqrt{-g} = d^2\zeta d^2\rho\sqrt{-\tilde{g}}$, where $\tilde{g}_{\mu\nu}(\tilde{\zeta})$ is the induced metric with respect to the coordinates $\tilde{\zeta}^{\mu} = \left\{\zeta^{a},\rho^{\alpha}\right\}$. To zeroth order in $\rho^{\alpha}$, this metric is given by a block diagonal matrix, such that $\tilde{g}_{\mu\nu} =$ diag$(\gamma_{ab},\delta_{\alpha\beta}) + \mathcal{O}(\rho)$, where $\delta_{\alpha\beta}$ is the usual Kronecker delta symbol and $\gamma_{ab}$ is the induced metric on the world-sheet defined by $\langle \phi \rangle = 0$, $\gamma_{ab}(\zeta) = g_{\mu\nu}(X)(\partial X^{\mu}/\partial \zeta^{a})(\partial X^{\mu}/\partial \zeta^{b})$. Substituting from (\ref{NO_ansatz}) and (\ref{NO_f}) and performing the transverse integration over $d^2\rho$, then gives \cite{CS_EFF.ACT}
\begin{eqnarray} \label{NG_act}
S = -\mu \int d^2\zeta \sqrt{-\gamma} + \ . \ . \  .
\end{eqnarray}
Hence, to leading order, the effective action for a Nielsen-Olesen (vacuum) string, in the Abelian-Higgs model,  is simply the Nambu-Goto action for an $F-$ string with no electric or magnetic world-sheet fluxes \cite{Nambu-Goto}.

\subsection{Microphysical ansatz and effective action for a superconducting string} \label{Sect3.2}

Next, we consider how a string becomes superconducting. The first concrete model of a superconducting string, carrying a bosonic current, was proposed by Witten, who considered the ``chiral" action \cite{Witten85}
\begin{eqnarray} \label{chiral_act}
S &=& \int \sqrt{-g}d^4x\bigg\{D_{\mu}\phi D^{\mu*}\phi^{*}  - \frac{1}{4}F_{\mu\nu}F^{\mu\nu}
\nonumber\\
&+& \tilde{D}_{\mu}\tilde{\phi} \tilde{D}^{\mu*}\tilde{\phi}^{*} - \frac{1}{4}\tilde{F}_{\mu\nu}\tilde{F}^{\mu\nu} - V(|\phi|, |\tilde{\phi}|) \bigg\}
\end{eqnarray}
where
\begin{eqnarray} 
V(|\phi|, |\tilde{\phi}|) = \frac{\lambda}{4}(|\phi|^2 - \eta^2)^2 + \frac{\tilde{\lambda}}{4}(|\tilde{\phi}|^2 - \tilde{\eta}^2)^2 + \tilde{\beta}|\phi|^2 |\tilde{\phi}|^2.
\nonumber
\end{eqnarray}
\begin{eqnarray} \label{AH_poy}
{}
\end{eqnarray}
Clearly, the action (\ref{chiral_act}) is simply a ``double copy" of the Abelian-Higgs action, including an appropriate interaction term between the two scalar fields, where the parameter $\tilde{\beta}$ is the coupling constant. It is therefore invariant under local $U(1) \times \tilde{U}(1)$ gauge transformations.

In order for a bosonic superconducting string to form, the $\tilde{U}(1)$ symmetry must be broken, giving rise to a vacuum string described by the ansatz (\ref{NO_ansatz}), whereas the additional $U(1)$ symmetry must remain unbroken. This requires $\tilde{\lambda}\tilde{\eta}^4 > \lambda\eta^4$ \cite{VS_book,Witten85}. Under these conditions, it is energetically unfavorable for $\tilde{\phi}$ to vanish everywhere, so that, although it vanishes outside the string, its expectation value within the core region is of order $(\tilde{\beta}\eta^2/\lambda)^{1/2}$. A superconducting string is then described by the microphysical ansatz
\begin{eqnarray} \label{chiral_ansatz}
\phi(r,\theta,z,t) = \eta f(r)e^{i\Gamma(z,t)},
\nonumber\\
A_0 = A_0(z,t),  \ A_z = A_0(z,t),
\end{eqnarray}
together with $A_{\theta} = 0$, $A_r = 0$, in addition to an ansatz of the form (\ref{NO_ansatz}) for the $\tilde{U}(1)$ sector (i.e. (\ref{NO_ansatz}) holds under the transformations $f(r) \rightarrow \tilde{f}(r)$, $a_{\theta}(r) \rightarrow \tilde{a}_{\theta}(r)$, etc.). Here, $\Gamma(z,t)$ is an arbitrary function describing ``twists" in the lines of constant phase of the $U(1)$ scalar field, along the length of the string. The boundary conditions for $f(r)$ are the inverse of those for $\tilde{f}(r)$, i.e.
\begin{eqnarray} \label{chiral_f_bc}
f(r) = \left \lbrace
\begin{array}{rl}
1,& \ r=0 \\
0,& \ r \rightarrow \infty,
\end{array}\right.
\end{eqnarray}
and $A_0(z,t)$, $A_z(z,t)$ are defined to be nonzero only within the string core \cite{Witten85}. This contributes an additional part to the effective action (\ref{NG_act}), given by
\begin{eqnarray} \label{chiral_DeltaS-2}
\Delta S = \frac{1}{4e^2\Sigma}\int d^2\zeta \sqrt{-\gamma}\gamma_{ab} J^{a}J^{b}
\end{eqnarray}
where $J^{a}$ denotes the world-sheet components of the physical current, $J^{a} = -\delta (\Delta S)/\delta A_{a} = -2e\Sigma(\partial^{a}\Gamma + eA^{a})$, and $\Sigma = 2\pi \eta^2\int rdr |f|^2$, which is finite since $\phi$ vanishes outside a width $\sim \eta^{-1}$ \cite{Witten85}. The effective action may then be written as
\begin{eqnarray} \label{chiral_S}
S = -\mu  \int d^2\zeta \sqrt{-\gamma}(1 - \gamma_{ab} j^{a}j^{b}),
\end{eqnarray}
where
\begin{eqnarray} \label{chiral_j}
j^{a} = (4e^2\Sigma\mu)^{-1/2}J^{a}
\end{eqnarray}
is the dimensionless current. The world-sheet energy momentum tensor for the superconducting string is defined as $\theta^{ab} = 2j^{a}j^{b} - \gamma^{ab}(\gamma_{cd}j^{c}j^{d})$, so that the physical energy-momentum tensor is given by \cite{CS_EFF.ACT}
\begin{eqnarray} 
T^{\mu\nu} \equiv \frac{-2}{\sqrt{-g}}\int d^2\zeta \sqrt{-\gamma}(\gamma^{ab} + \theta^{ab})\partial_{a}X^{\mu}\partial_{b}X^{\nu}\delta^{4}(x-X)
\nonumber
\end{eqnarray}
\begin{eqnarray} \label{chiral_EMT}
{}
\end{eqnarray}
where $X^{\mu}$ denotes a string embedding coordinate, as before, and $x^{\mu}$ denotes a space-time background coordinate.

In \cite{Witten85}, Witten also outlined, from a microphysical perspective, how strings can carry fermionic currents, with fermionic charge carriers trapped as zero modes along the string. However, due to Bose-Fermi equivalence in $(1+1)$ dimensions, the effective action (\ref{chiral_S}) developed by CTH is valid for strings with both bosonic and fermionic currents \cite{CS_EFF.ACT}. It is therefore valid more generally, as a first order approximation, for any species of superconducting string.

For string loops, it is self-evident that the current persists in time due the existence of a second topological invariant $N \in \mathbb{Z}$, given by
\begin{eqnarray} \label{chiral_N}
N = \frac{1}{2\pi}\int_{0}^{2\pi}\frac{\partial \Theta}{\partial \sigma} d\sigma,
\end{eqnarray}
where $\sigma \in [0,2\pi)$ is a space-like parameter along the string length. Thus, $|N|$ gives the number of twists in the phase $\Theta$ of $\phi$ within the loop. This follows directly from the imposition of periodic boundary conditions to ensure continuity.

\subsection{Constants of motion for long, straight, superconducting strings} \label{Sec.3.3}

For long strings we may consider finite sections of length $L$, containing $N$ twists, for which
\begin{eqnarray} \label{L-1}
L = \lambda_z N
\end{eqnarray}
for some characteristic length scale $\lambda_z$. In the case of constant currents, $\Gamma(z,t)$ is linear in both $z$ and $t$ -- or, equivalently, in $\sigma$ and $\tau$, or $\sigma$ and $t$, where $\sigma$ and $\tau$ are dimensionless space-like and time-like world-sheet parameters, respectively -- such that
\begin{eqnarray} \label{Gamma-1}
\Gamma(z,t) = k_z z + \omega_z t \equiv \Gamma(\sigma,t) = N\sigma + \omega_z t,
\end{eqnarray}
where
\begin{eqnarray} \label{Gamma-2}
\omega_z = \pm k_z, \ \ \ k_z \equiv \frac{2\pi}{\lambda_z} = \frac{2\pi N}{L}
\end{eqnarray}
and $\lambda_z$ represents the (constant) distance over which a single twist in the phase of $\phi$ occurs. Even if nonlinear oscillations in the phase are present, corresponding to fluctuations in the current, we may simply replace $\lambda_z$ with $\langle \lambda_z \rangle$, the spatially averaged value, which, by current conservation, must also be independent of $t$.

Taking the embedding for a finite section $L$ of long, straight string
\begin{eqnarray} \label{long_string_emb}
X^{\mu}(\sigma,\tau) = (t(\tau) = \xi \tau, x=0, y=0, z(\sigma) = (2\pi)^{-1}L\sigma) \
\end{eqnarray}
where $\xi$ is a constant with dimensions of time, we have
\begin{eqnarray} \label{null_current-1}
\mathcal{J}^2 \equiv \gamma_{cd}j^{c}j^{d} = (j^{0})^2 - (j^{z})^2 = 0,
\end{eqnarray}
for null currents, where we have defined
\begin{eqnarray} \label{null_current-2}
j^{0} \equiv \xi j^{\tau}, \ \ \ j^{z} \equiv (2\pi)^{-1}L j^{\sigma}.
\end{eqnarray}
We note that the null current condition implies
\begin{eqnarray} \label{null_current-3}
j^{0} = \pm j^{z}.
\end{eqnarray}
The constants of motion are given by
\begin{eqnarray} \label{CoM-1}
\Pi^{\mu} = \int k^{(\mu)}{}_{\nu}T^{0\nu}\sqrt{-g}d^3x,
\end{eqnarray}
where $\left\{k^{(\mu)}{}_{\nu}\right\}$ are the set of space-time Killing vectors. To help clarify the physical picture, we choose to denote the energy $\Pi^{0} = E$ and to label $\Pi^{i} = P^{i}$, if $x^{i}$ corresponds to a Cartesian coordinate, and $\Pi^{j} = l^{j}$ if $x^{j}$ corresponds to an angular coordinate. Thus, components of linear momentum are denoted by the letter $P$ and components of angular momentum by the letter $l$. Though we need not consider components of angular momentum in this section, the distinction will become important in Sec.~\ref{Sect5}, when we deal with higher-dimensional strings.

For the sake of simplicity, we assume that the string carries constant current, $j^{0} = \pm j^{z} = {\rm const.}$ It is then straightforward to show that the energy, momentum and integrated tension of a section of superconducting string of length $L$ are, respectively \cite{CS_EFF.ACT}
\begin{eqnarray} \label{chiral_E}
E = (1 + j^2)\mu L, \
T^{z} = -(1 - j^2)\mu L, \
P^{z} = \pm j^2\mu L,
\end{eqnarray}
where we have again assumed critical coupling, $\lambda = 2e^2$, and where we have defined
\begin{eqnarray} \label{chiral_j^2}
j^2 \equiv \frac{1}{2}(j^{0})^2 = \frac{1}{2}(j^{z})^2,
\end{eqnarray}
for convenience. This must not be confused with $\mathcal{J}^2 \equiv \gamma_{ab}j^{a}j^{b}$, which is zero for null currents (\ref{null_current-1}). The conserved electric charge is
\begin{eqnarray} \label{chiral_Q}
Q = \pm  e j\mu L r_c,
\end{eqnarray}
and the choice of sign in Eqs. (\ref{chiral_E}) and (\ref{chiral_Q}) must match that taken in (\ref{null_current-3}). Even in the case of $t-$ and $z-$dependent oscillations in the current, we may simply replace $j^2$ with $\langle j^2 \rangle$, the spatially averaged value, which is also constant in time, by virtue of current conservation. In the more general scenario, expressions of the form (\ref{chiral_E}) and (\ref{chiral_Q}) remain valid under the transformations $j^2 \rightarrow \langle j^2 \rangle$ and $j \rightarrow \langle j \rangle$, respectively. In this limit, we adopt a phenomenological description of the superconducting string, based on obtaining the effective energy-momentum tensor in the wire approximation \cite{Anderson_book,VS_book}.

Finally, we note that the threshold current (\ref{Jmax}) may be written as $J_{\rm max} \approx e/r_c$ for the models considered above. For bosonic currents, exceeding this threshold implies electric field strengths large enough to induce pair production via the Schwinger process whereas, for trapped fermions, this marks the critical point at which it becomes energetically favorable for particles to leave the string \cite{VS_book}. In our phenomenological model, this limit implies the existence of a minimum wavelength associated with the current, $\lambda_z^{\rm min} = 2\pi r_c$. This will be shown explicitly in Sec. \ref{Sect5} by considering the higher-dimensional string case, where $\lambda_z$ may be interpreted as the distance over which the string wraps a single winding in the compact space and the string thickness $r_c$ is formally equivalent to the compactification radius $R$. Due to the equivalence of higher-dimensional strings (under dimensional reduction) and superconducting strings in four dimensions \cite{dim_red}, it therefore remains valid, even for field-theoretic strings described by the effective action (\ref{chiral_act}). This implies the existence of a maximum winding number, within the length $L$, given by $N_{\rm max} = L/(2\pi r_c)$.

\section{Mass transfer to a black hole pierced by two antipodal superconducting strings with opposite charge and current flow} \label{Sect4}

\subsection{Initial estimates for topological defect strings} \label{Sect4.1}

The energy $\Delta E$, in addition to the rest mass energy $E =  \mu L$, due to the flow of charge within the length $L$, is
\begin{eqnarray} \label{dE}
\Delta E = j^2 \mu c^2 L = c\Delta P,
\end{eqnarray}
where $\Delta P \equiv P^z$ is the additional momentum, as expected for the flow of a relativistic fluid. (For the sake of clarity, the physical constants $c$, $G$ and $\hbar$ and are explicitly included in all expressions throughout the reminder of this section.) In time $t$, an infinitesimal volume of charged fluid covers a distance $L(t) = ct$, so that the excess energy flowing into the black hole from two antipodal strings is
\begin{eqnarray} \label{dE(t)}
\Delta_{2} E(t) = 2j^2 \mu c^3 t.
\end{eqnarray}
The total energy of the black hole increases linearly with time but the exact rate is controlled by the values of the model parameters $\mu$ and $j^2$. If the strings carry opposite charge but equal current (flowing in opposite directions), the net charge and momentum transfer to the black hole is zero. Physically, in our effective model, it is not simply mass that flows into the black hole but also kinetic energy associated with its flow. Since the net momentum gain of the black hole is zero, this energy is converted into rest mass when it enters the event horizon.

Extending the model to include $\mathcal{N} \in \mathbb{N}$ antipodal pairs is straightforward: the net charge and momentum transfer are still zero and $\Delta_{2\mathcal{N}} E(t) = \mathcal{N}\Delta_{2}E(t)$. However, for small seed black holes, it is unrealistic to expect large numbers of strings to pierce the horizon simultaneously. Nonetheless, as the black hole accretes from the string network and grows, it may (in principle) intersect with multiple strings, depending on the average string density at that epoch. Alternatively, multiple strings attached to a single black hole may be formed in a system such as that described in \cite{Ary86}, in which charged matter also becomes trapped in the strings during the process of formation via the usual Kibble mechanism \cite{VS_book}. Although these possibilities are interesting, their thorough investigation lies outside the scope of the present work and, for now, we note that the estimates of mass accretion given in the remainder of this work are valid, to within an order of magnitude, for $\mathcal{N} \sim \mathcal{O}(1)$. Further remarks on mechanisms by which the seed black holes may come to intersect with the string network are given in Sec. \ref{Sect6}.

In principle, we may consider arbitrarily large currents in the classical theory. However, as discussed in Secs. \ref{Sect2} and \ref{Sect3.2}, both bosonic and fermionic superconducting strings are expected have critical maximum currents of order $J_{\rm max} \approx qc/r_c$, for a given fundamental charge $q$ and string width $r_c$, due to quantum effects. (We here assume that the scalar and vector core radii of a defect string are of the same order of magnitude, though, even if this assumption is relaxed, the threshold current may be expressed in terms of a single effective core radius ``$r_c$" that depends on both $r_s$ and $r_v$.) Assuming $r_c \geq l_{Pl}$, where $l_{Pl} = \sqrt{\hbar G/c^3} = 1.617 \times 10^{-35}$ m is the Planck length, we may parameterize $r_c$ as some multiple of its minimum possible value. Likewise, assuming $q \leq q_{Pl}$, where $q_{Pl} = \sqrt{4\pi \epsilon_0 \hbar c} = 1.876 \times 10^{-18}$ C is the Planck charge, we may parameterize any current $J$ in terms of the maximum possible threshold current, the ``Planck current", $J_{Pl} = q_{Pl}c/l_{Pl} = 3.479 \times 10^{25}$ A. Hence, we may write
\begin{eqnarray} \label{JR_limits}
r_c = \beta l_{Pl}, \  (\beta \geq 1); \ \ \ J =  \frac{q_{Pl} c}{\gamma l_{Pl}}, \ (\gamma \geq 1)
\end{eqnarray}

For a general string species, we may express the dimensionless current as $j = J/J_{\rm max}$, or, equivalently,
\begin{eqnarray} \label{JR_limits*}
j = \left(\frac{qc}{r_c}\right)^{-1}J = \frac{\beta}{\sqrt{\alpha_q}\gamma},
\end{eqnarray}
where $\alpha_q = q^2/q_{Pl}^2 \leq 1$. This follows from the fact that the current threshold $J = J_{\rm max}$ coincides with the limit in which the effective (classical) string tension $\mathcal{T}^{z}$ undergoes a transition to become an effective pressure. (If permitted, this would effectively turn the string into a repulsive rod \cite{rep_rod}.) This will be shown explicitly in Sec. \ref{Sect5}, by means of the formal equivalence between superconducting strings and higher-dimensional strings under dimensional reduction \cite{dim_red}. For now, we note that this statement holds true for Witten's chiral string model, since $\Sigma \sim 1/\lambda$ and $\mu \sim \tilde{\eta}^2$ \cite{Witten85}, so that $j \sim Jr_c/q$, where $r_c \sim \sqrt{\lambda}\tilde{\eta}^{-1}$. In the limit $q \rightarrow q_{Pl}$, $r_c \rightarrow l_{Pl}$ we have $j^2 = \gamma^{-2}$. In this case, the limit $j^2 \rightarrow 1$ is reached when $\gamma \rightarrow 1$, corresponding to $J \rightarrow J_{Pl}$. However, in general, the following condition must hold:
\begin{eqnarray} \label{j^2_limits}
j^2 \leq 1 \iff \gamma^2 \geq \beta^2/\alpha_q.
\end{eqnarray}
Generically, saturation of the bound (\ref{j^2_limits}) occurs for $\gamma = \beta/\sqrt{\alpha_q}$ and, in principle, this may be realized even in models for which $q < q_{Pl}$, $r_c > l_{Pl}$, $J_{\rm max} < J_{Pl}$.

Since, generically, we may also assume that the mass per unit length of the string is related to the core radius via
\begin{eqnarray} \label{T}
\mu = \frac{\kappa^2 \hbar }{(2\pi)^2r_c^2 c}, \ (\kappa^2 \geq 1)
\end{eqnarray}
the total mass transfer to the black hole  in time $t$ is given by
\begin{eqnarray} \label{15}
\Delta_2 M(t) = \frac{1}{2\pi^2\alpha_q}\left(\frac{\kappa}{\gamma}\right)^2\left(\frac{G}{c^2}\right)^{-1}ct.
\end{eqnarray}
Assuming that the seed mass is negligible, that the string$-$black hole system formed at, or soon after, the end of the inflationary epoch (i.e. at approximately $t=0$ on a cosmological time scale), and that energy input from two antipodal strings is the dominant accretion mechanism, this implies that black holes of mass $M = 2 \times 10^9 M_{\odot}$ are able to form within a time interval
\begin{eqnarray} \label{dt}
\Delta_2 t = 8\pi^2 \times 10^9 \alpha_q \frac{G}{c^3}\left(\frac{\kappa}{\gamma}\right)^{-2} M_{\odot}.
\end{eqnarray}
In order for $\Delta_2t \leq 2.43 \times 10^{16}$ s $= 4.51 \times 10^{59} t_{Pl}$, where $t_{Pl}=l_{Pl}/c$ is the Planck time, this requires
\begin{eqnarray} \label{bound}
\gamma^2 \leq 5.708 \times 10^{48}\left(\frac{m_{Pl}}{M_{\odot}}\right)\frac{\kappa^2}{\alpha_q} = 6.213 \times 10^{10}\frac{\kappa^2}{\alpha_q}.
\end{eqnarray}
where $m_{Pl} = \sqrt{\hbar c/G} = 2.176 \times 10^{-5}$ g is the Planck mass.

Next, we assume, for simplicity, that the expression for the intrinsic string tension takes a form analogous to that given in the chiral string model at critical coupling, i.e.
\begin{eqnarray} \label{T-2}
\mu = 2\pi \tilde{\eta}^2 |n|\left(\frac{G}{c^2}\right)^{-1},
\end{eqnarray}
where $\tilde{\eta} \times m_{Pl}c^2$ is the symmetry breaking energy scale at which string formation occurred and $n$ is the topological winding number. In this case, we may also assume that the expression for the string width takes the general form \cite{VS_book}
\begin{eqnarray} \label{}
r_c \approx (\sqrt{\alpha_q} \tilde{\eta})^{-1}l_{Pl}.
\end{eqnarray}
Comparing this with (\ref{JR_limits}) we have $\beta = (\sqrt{\alpha_q}\tilde{\eta})^{-1}$ and comparing (\ref{T-2}) with (\ref{T}) gives $\kappa^2 = (2\pi)^3\tilde{\eta}^2|n|\beta^2$, so that
\begin{eqnarray} \label{kappa}
\kappa^2 = (2\pi)^3|n|/\alpha_q.
\end{eqnarray}
Combining (\ref{kappa}) with (\ref{j^2_limits}) and (\ref{bound}) then gives
\begin{eqnarray} \label{gamma_bounds}
\alpha_q^{-2}\tilde{\eta}^{-2} \leq \gamma^2 \leq 1.541 \times 10^{13}|n|\alpha_q^{-2}.
\end{eqnarray}
The limits in (\ref{gamma_bounds}) correspond to physical currents in the range
\begin{eqnarray} \label{J_bounds}
\beta^{-1}J_{Pl} \geq J \geq 2.547 \times 10^{-7} \frac{\alpha_q}{\sqrt{|n|}}J_{Pl},
\end{eqnarray}
and to dimensionless currents
\begin{eqnarray} \label{j_bounds}
1 \geq j \geq 2.457 \times 10^{-7}(\sqrt{|n|}\tilde{\eta})^{-1}.
\end{eqnarray}

Adopting the most conservative estimate for the topological winding number $|n| \sim \mathcal{O}(1)$ and considering GUT scale strings, for which $\eta^2_{\rm GUT} = m_{\rm GUT}^2/m_{Pl}^2 \approx 6.703 \times 10^{-7}$, whose existence is marginally consistent with current observational bounds on the intrinsic string tension \cite{string_tens_bounds}, then implies
\begin{eqnarray} \label{j_GUT}
j^2_{\rm GUT} \geq 9.679 \times 10^{-8} \approx 10^{-7}.
\end{eqnarray}
This gives the minimum value of the square of the dimensionless current required for GUT scale strings, in order for current transfer from the string network to a seed black hole to produce an SMBH of mass $2 \times 10^9 M_{\odot}$ by $z = 7.085$. For $q=e$ ($\alpha_e \approx 1/137$), this corresponds to a physical current of order
\begin{eqnarray} \label{J_GUT}
J_{\rm GUT} \geq 6.463 \times 10^{16} \ {\rm A},
\end{eqnarray}
well within the threshold current estimated by Witten \cite{Witten85}. It is straightforward to check that both the ``simplistic'' analysis presented in Sec. \ref{Sect2} and the effective action approach presented here give the same (maximum) mass increase, $M_{\rm max}(t) \sim (m_q^2c^2/\hbar)t$ for $J_{\rm max} \sim qm_qc^2/\hbar$. Our previous statements regarding the necessary efficiency of accretion from GUT scale superconducting strings therefore remain valid.

\subsection{Refined analysis for topological defect strings} \label{Sect4.2}

Though the analysis above provides a rough estimate for the mass accretion rate of black holes connected to superconducting strings, due to the influx of current, there is an additional subdominant mechanism that will further enhance this process. As current flows into the black hole, increasing its mass linearly in time, its horizon radius will also grow linearly in time. The black hole will, in turn, ``swallow" an increasingly large section of string, absorbing both the kinetic energy associated with the current and the rest mass of the string itself. The energy gain due to current absorption in time $t$, Eq. (\ref{dE(t)}), causes the Schwarzschild radius to expand according to
\begin{eqnarray} \label{}
\Delta_2 R_S(t) = \frac{2G}{c^2} \Delta_{2} M(t) =  4j^2\mu \left(\frac{G}{c^2}\right)ct.
\end{eqnarray}
This implies an additional mass accretion
\begin{eqnarray} \label{}
\Delta_2 M'(t) = 4j^2(1 + j^2)\mu^2\left(\frac{G}{c^2}\right)ct.
\end{eqnarray}
The total mass gain, in time $t$, is therefore
\begin{eqnarray} \label{}
\Delta_2 M_{\rm tot}(t) &\approx &2\mu j^2 \left[1 + 2\mu(1 + j^2)\frac{G}{c^2}\right] ct.\nonumber\\
\end{eqnarray}
Imposing the limit $j^2=1$, this gives a correction to our previous estimates of order $\mathcal{O}(G\mu/c^2)$. For GUT scale strings this is equivalent to $\eta_{\rm GUT}^2 \sim \mathcal{O}(10^{-7})$ and may neglected. At most, for strings with widths/energies comparable to the Planck scales, the additional contribution will be of order unity, so that the estimates given above remain valid, to within an order of magnitude.

\section{Accretion rates in ``exotic" higher-dimensional scenarios} \label{Sect5}

In this Section, we review the basic properties of a section of long straight string containing windings in a compact internal space. It is shown that, under dimensional reduction, the motion of the string in the compact space gives rise to an effective world-sheet flux, from a four-dimensional perspective, which is formally analogous to the superconducting current parameter $j^2$ in the CTH action \cite{CS_EFF.ACT}. For the constant current case, we determine $j^2$ in terms of higher-dimensional variables (including the compactification radius of the extra dimensions $R$, and winding number, $m$) and the fundamental string tension, $\mathcal{T}$. We then conduct a similar analysis to that given in Sec. \ref{Sect4}, thus obtaining bounds on the region of higher-dimensional parameter space compatible with SMBH formation at $z \gtrsim 7$.

\subsection{The higher-dimensional $F-$string} \label{Sect5.1}

Beginning with the Nambu-Goto action for an $F-$string with fundamental tension $\mathcal{T}$, carrying no additional world-sheet fluxes,
\begin{eqnarray} \label{NG_act*}
S = -\mathcal{T} \int d^2\zeta \sqrt{-\gamma},
\end{eqnarray}
we consider the string embedding
\begin{eqnarray} \label{HD_emb}
X^I(\sigma,\tau)=\big(t(\tau) = \xi\tau, x = 0, y = 0,
\nonumber\\
z(\sigma) = (2\pi)^{-1}L \sigma, \psi(\sigma,\tau)\big)
\end{eqnarray}
where $\psi \in [0,2\pi)$ denotes an angular coordinate in the compact internal space. The string tension is given by
\begin{eqnarray} \label{F-string_tens}
\mathcal{T} = (2\pi \alpha')^{-1}.
\end{eqnarray}
where $\alpha'$ denotes the Regge slope parameter, which is related to the fundamental string length scale via $l_{\rm str} = \sqrt{\alpha'}$ \cite{superstrings}. (Throughout this subsection, we again set $c = \hbar = G =1$ for convenience.) Strictly, the ansatz (\ref{HD_emb}) describes a section of long straight string, with four-dimensional length $L$, embedded in a higher-dimensional spacetime with a single compactified direction, so that we may assume a metric of the form
\begin{eqnarray} \label{HD_metric}
ds^2 = dt^2 - dr^2 - r^2b^2d\theta^2 - dz^2 - R^2d\psi^2.
\end{eqnarray}
Here, $b^2$ is again related to the angular deficit induced by the presence of the string, via $b^2 = 1- \delta$, and $R$ is the compactification radius of the extra dimension. This corresponds to an embedding in $\tilde{M}^4 \times S^1$ where $\tilde{M}^4$ denotes the usual four-dimensional Minkowski space with an angular deficit $\delta$.

However, since strings are one-dimensional objects, they necessarily wrap (topologically) $S^1$ subcycles in any internal space. As such, (\ref{HD_metric}) may be taken as an effective metric for any embedding with constant winding radius. The following results are therefore valid more generally, for wound strings in \emph{almost} arbitrary internal spaces $-$ the only caveat being that at least one $S^1$ subcycle of constant radius must exist. (For example, in \cite{WS_KS}, strings wrapping great circles on the $S^3$ internal space that regularizes the warped deformed conifold of the Klebanov-Strassler geometry \cite{Klebanov:2000hb}, a toy model for a realistic higher-dimensional cosmology obtained via a flux-compactification in string theory \cite{Denef:2007pq}, were considered.) Clearly, (\ref{HD_emb}) is equivalent to the long string embedding (\ref{long_string_emb}), except for the presence of the higher-dimensional part.

For this embedding, the constants of motion and effective integrated tension of the $F-$string may be written as:
\begin{eqnarray} \label{HD_E}
E &=& \mathcal{T} L\Omega^{-2}, \ \ \ \ \ \ \ \ \ \ \ \ \ T^{z} = -\mathcal{T} L\left(\frac{2\Omega^{2}-1}{\Omega^{2}}\right),
\nonumber\\
P^{z} &=& \pm \mathcal{T} L \left(\frac{1-\Omega^{2}}{\Omega^{2}}\right), \ l^{\psi} = \pm \mathcal{T} L \frac{\sqrt{1-\Omega^{2}}}{\Omega},
\end{eqnarray}
where the function $\Omega^2(t)$ is defined as
\begin{eqnarray} \label{Omega^2}
\Omega^{-2}(t) \equiv \frac{1}{2\pi}\int_{0}^{2\pi}\frac{L^2 + (2\pi)^2R^2(\partial_{\sigma} \psi)^2}{L^2} d\sigma.
\end{eqnarray}
Physically, $\Omega^2(t)$ represents the fraction of the total string length lying in the large dimensions and $1-\Omega^2(t)$ represents the fraction contained in the compact space. In general, this may be time-dependent -- for example, for oscillating loops for which $L=L(t)$ -- but, for long straight strings, $\Omega^2 = {\rm const.}$ This follows directly from the imposition of periodic boundary conditions on $\psi$, such that $\psi(m\sigma,t) = \psi(m(\sigma + 2\pi p),t)$ for all $p \in \mathbb{Z}$, which is necessary for the conservation of energy and momentum within $L$. In our chosen coordinate system, the string EOM in $\psi(\sigma,\tau) \equiv \psi(z,t)$ gives
\begin{eqnarray} \label{HD_string_EOM}
\dot{\psi}^2 = \psi'^2, \ \ \ \dot{\psi} = \pm \psi',
\end{eqnarray}
where a dot denotes differentiation with respect to $t$ and a dash denotes differentiation with respect to $z$, respectively, and the choice of sign in (\ref{HD_string_EOM}) determines the sign of $l^{\psi}$ in (\ref{HD_E}).

We see immediately, by comparison of Eqs. (\ref{chiral_E}) and (\ref{HD_E}), the following equivalence between four-dimensional and higher-dimensional parameters:
\begin{eqnarray} \label{HD_j^2}
\langle j^2 \rangle = \left(\frac{1-\Omega^{2}}{\Omega^{2}}\right), \ \ \ \sqrt{\langle j^2 \rangle} = \pm \frac{\sqrt{1-\Omega^{2}}}{\Omega}
\end{eqnarray}
(First consider replacing $j^2$ with $\langle j^2 \rangle$ in Eqs. (\ref{chiral_E}), in order to obtain the more general expression valid for non-constant current.) In other words, we see that, under dimensional reduction, the motion of the windings in the compact space gives rise to an effective four-dimensional world-sheet current and there is a formal correspondence between the current density and the momentum in the extra dimensions, as originally shown by Nielsen \cite{dim_red}. (See also \cite{string_4D_eff.} for a more detailed treatment of the effective four-dimensional dynamics of strings with higher-dimensional windings, and their stability.)

The constant current case corresponds to the existence of linear windings in the higher-dimensional space, such that
\begin{eqnarray} \label{linear_psi}
\psi(z,t) = k_{z}z +  \omega_z t \equiv \psi(\sigma,t) = m\sigma + \omega_z t,
\end{eqnarray}
$m \in \mathbb{Z}$, and the string EOM (\ref{HD_string_EOM}) implies the existence of a dispersion relation of the form
\begin{eqnarray} \label{HD_disp_rel}
\omega_z = \pm k_z, \ \ \ k_z \equiv \frac{2\pi}{\lambda_z} = \frac{2\pi m}{L},
\end{eqnarray}
where
\begin{eqnarray} \label{HD_L-1}
L = \lambda_z m.
\end{eqnarray}
These expressions may be compared with Eqs. (\ref{L-1})-(\ref{Gamma-2}). Hence, the winding number
\begin{eqnarray} \label{Omega^2}
m = \frac{1}{2\pi}\int_{0}^{2\pi}\frac{\partial \psi}{\partial \sigma} d\sigma
\end{eqnarray}
plays \emph{almost} the same role as Witten's topological invariant $N$ but, here, a caveat is necessary, since $m$ is only a topological invariant if the compact space is not simply connected. For simply connected internal manifolds, $m$ must be stabilized, if at all, dynamically (see \cite{WS_KS,string_4D_eff.} for examples). Nonetheless, in either scenario there exists a formal correspondence between the higher-dimensional string embedding $\psi(z,t)$ (\ref{linear_psi}) and the function $\Gamma(z,t)$ (\ref{Gamma-1}) that determines the embedding of the lines of constant phase for the $U(1)$ scalar field of the chiral string. Hence, we have
\begin{eqnarray} \label{HD_const_current}
j^2 = \frac{(2\pi)^2R^2}{\lambda_z^2} = {\rm const.}
\end{eqnarray}
so that
\begin{eqnarray} \label{HD_j^2=1}
j^2 = 1 \iff \Omega^2 = 1/2,
\end{eqnarray}
or equivalently $\lambda_z = \lambda_z^{\rm min}$ where
\begin{eqnarray} \label{lambda_z^{min}}
\lambda_z^{\rm min} = 2\pi R,
\end{eqnarray}
as claimed in Sec. \ref{Sect3.2}. In this limit the string becomes effectively tensionless from a four-dimensional perspective, though its intrinsic tension remains unchanged \cite{WS_KS,string_4D_eff.}.

For circular loops, this scenario corresponds to the existence of cosmic vortons \cite{vortons}. The closer superconducting string loops come to a vorton-type configuration, the less influence we expect small-scale kinks and propagating ``wiggles" to have on their evolution and corresponding lifetimes. In the ``pure" vorton case, in which string loops may be extremely long-lived, potentially leading to a ``vorton excess problem" \cite{vortons}, it is interesting to note that the intersection of seed black holes with large numbers of small vorton loops would help to alleviate this problem. (In the higher-dimensional wound-string case, considered in Sec. IVB, the equivalent problem would be the ``cycloops excess", considered by Avgoustidis and Shellard \cite{Avgoustidis:2005vm}). That said, we do not claim a potential solution to this problem via our proposed mechanism, since this would necessarily require a large number of string--black hole intersections. We note, however, that such a scenario may be viable if there existed of a network of ``cosmic necklaces" -- whether of string theory, or field theoretic-origin -- in the very early Universe \cite{Lake:2010qsa,CS_NECKLACE}.

For the sake of completeness, we now  consider the analogue of electric charge in the higher-dimensional $F-$string model, which is related to the conservation of angular momentum in the compact space. It is useful to first define quantity
\begin{eqnarray} \label{lambda_psi}
\Lambda^{\psi} = \pm \sqrt{-l^{\psi}l_{\psi}}
\end{eqnarray}
where $l_{\psi} = -R^2l^{\psi}$ and the sign of $\Lambda^{\psi}$ is chosen to match that of $l^{\psi}$. Thus,
\begin{eqnarray} \label{HD_l^psi}
\Lambda^{\psi} = \pm \mathcal{T} R L \frac{\sqrt{1-\Omega^{2}}}{\Omega}.
\end{eqnarray}
Under dimensional reduction $q\Lambda^{\psi}/L$ may be interpreted as a (null) conserved electric current
\begin{eqnarray} \label{HD_J}
J^{0} = \pm J^{z} = \pm q \mathcal{T} R \frac{\sqrt{1-\Omega^{2}}}{\Omega},
\end{eqnarray}
where $q$ is an arbitrary coupling, corresponding to the arbitrary constant multiplying a general Noether charge. In this model, we interpret it as the fundamental charge of a charge carrier. Current conservation then follows from the EOM for $\psi$ (\ref{HD_string_EOM}), which ensures conservation of $\Lambda^{\psi}$, and the conserved charge within the string length $L$ is
\begin{eqnarray} \label{HD_Q}
Q = \pm q \mathcal{T} R L \frac{\sqrt{1-\Omega^{2}}}{\Omega}.
\end{eqnarray}
Finally, we note that, in wound-string model, $R$ plays the role of an effective string thickness in four-dimensional space and appears in the same relative position in Eqs. (\ref{HD_E}) and (\ref{HD_Q}) as $r_c$ in Eqs. (\ref{chiral_E}) and (\ref{chiral_Q}).

\subsection{Momentum, charge and mass-energy transfer rate to a black hole pierced by two antipodal higher-dimensional strings with opposite windings} \label{Sect5.2}

\subsubsection{Initial estimates for higher-dimensional $F-$strings} \label{Sect5.2.1}

For the higher-dimensional string, the additional energy $\Delta E$, over and above the rest mass-energy corresponding to the length of string contained in the large dimensions, $E =  \mathcal{T}L$, is
\begin{eqnarray} \label{dE*}
\Delta E = \mathcal{T} \left(\frac{1-\Omega^2}{\Omega^2}\right)L = c\Delta P,
\end{eqnarray}
where $\Delta P = P^z$ is the additional momentum generated in the $z-$direction, due to the rotation of the windings in the compact space. Thus, the string behaves like a ``corkscrew", in which rotational motion in the extra dimensions is converted into linear motion along the four-dimensional string length \cite{WS_KS,string_4D_eff.}.

Since the effective current produced by this motion moves at the speed of light (\ref{HD_disp_rel}), in time $t$ an infinitesimal section of wound string covers a three-dimensional distance $L(t) = ct$. The excess energy flowing into the black hole from two antipodal strings, with windings rotating in opposite directions, is therefore
\begin{eqnarray} \label{dE(t)*}
\Delta_{2} E(t) = 2\mathcal{T} \left(\frac{1-\Omega^2}{\Omega^2}\right) ct =  2\mathcal{T} \frac{(2\pi)^2R^2}{\lambda_z^2} ct.
\end{eqnarray}
Again, the total energy of the black hole increases linearly with time but, in the higher-dimensional model, the exact rate is controlled by the values of the fundamental parameters $\mathcal{T}$ and $R$ together with the phenomenological parameter $\lambda_z$, the ``wavelength" of an individual winding that determines the current density. If the windings on each string wrap cycles in opposite directions, the strings carry opposite charge and, if the windings rotate in opposite directions, the currents flow in opposite directions. Hence, in our toy higher-dimensional model, the net charge and momentum transfer to the black hole is still zero.

In the classical theory, we may consider arbitrarily large values of $R$ and arbitrarily small values of $\lambda_z$, corresponding to arbitrarily large values of $j^2$ via Eq. (\ref{HD_j^2}). However, as shown previously in (\ref{HD_j^2=1}), the critical maximum current, due to quantum mechanical effects, corresponds to $j_{\rm max}^2=1$ or equivalently $\lambda_z^{\rm min} = 2\pi R$ (\ref{lambda_z^{min}}). In the chiral string picture, this is equivalent to requiring the distance between neighboring twists in the lines of constant phase to be greater than the circumference of the string. In the wound string picture, it implies that neighboring windings must be separated by a distance larger than the compactification scale.

In this case, we may assume $R \geq l_P$, and parameterize $R$ and $\lambda_z$ as multiples of their minimum possible values via
\begin{eqnarray} \label{*}
R = \beta l_{Pl}, \ \ \ \lambda_z = \gamma \lambda_z^{\rm min} = 2\pi \beta\gamma l_{Pl}, \ (\beta,\gamma \geq 1).
\end{eqnarray}
We may also assume
\begin{eqnarray} \label{T*}
\mathcal{T} = \frac{\kappa^2 \hbar c}{(2\pi)^2R^2}, \ (\kappa^2 \geq 1)
\end{eqnarray}
so that the total mass transfer to the black hole  in time $t$ is
\begin{eqnarray} \label{15*}
\Delta_2 M(t) = \frac{1}{2\pi^2}\left(\frac{\kappa}{\beta\gamma}\right)^2\left(\frac{G}{c^2}\right)^{-1}ct
\end{eqnarray}
As before, we assume that the seed mass is negligible, that the black hole formed at (or soon after) the end of the inflationary epoch, and that energy input from two antipodal strings is the dominant accretion mechanism. This implies that black holes with $M = 2 \times 10^9 M_{\odot}$, the mass of the SMBH powering quasar ULAS J112001.48+064124.3, are able to form within a time interval
\begin{eqnarray} \label{dt*}
\Delta_2 t = 3.911 \times 10^5 \left(\frac{\kappa}{\beta\gamma}\right)^{-2}\; {\rm s}.
\end{eqnarray}
Hence, $\Delta_2t \leq 0.77$ billion years requires
\begin{eqnarray} \label{bound*}
\gamma^2 \leq 6.213 \times 10^{10}\left(\frac{\kappa}{\beta}\right)^2.
\end{eqnarray}
Considering an $F-$string with intrinsic tension
\begin{eqnarray} \label{**}
\mathcal{T} = \frac{\hbar c}{2\pi l_{\rm str}^2},
\end{eqnarray}
where $l_{\rm str} = \hbar c\sqrt{\alpha'}$ \cite{superstrings}, we have $\kappa^2 = 2\pi \beta^2 (l_{Pl}/l_{\rm str})^2$, and the bound Eq. (\ref{bound*}) implies
\begin{eqnarray} \label{XX}
\gamma^2_{\rm str} \leq 3.904 \times 10^{11}(l_{Pl}/l_{\rm str})^2.
\end{eqnarray}

In the string theory picture, it is common to assume that the \emph{fundamental} string scale is of the order of the Planck scale, $l_{\rm str} \approx l_{Pl}$. However, Planck scale strings are clearly ruled out by current observational data  \cite{string_tens_bounds}. One solution to this problem is to take account of the fact that, since $F-$strings live in higher-dimensional spaces, the moduli of the internal space must be stabilized in some way. One way to stabilize the geometry of the compact space is via a flux-compactification, which indtroduces a phenomenological ``warp factor" $a^2 \in (0,1]$, multiplying the components of the four-dimensional part of the metric. This is induced by the back-reaction of the fluxes on the large dimensions \cite{Denef:2007pq}. (See also \cite{Klebanov:2000hb} as an example of a flux-compactified geometry and \cite{Lake:2010qsa,WS_KS,string_4D_eff.}, plus references therein, for applications of flux compactifications to cosmic string phenomenology.) In this scenario, the effective tension is reduced, giving $\tilde{\mathcal{T}} = a^2\mathcal{T} = \kappa^2\hbar c/((2\pi)^2R^2)$ in place of (\ref{T*}), so that $\kappa^2 = 2\pi a^2 \beta^2 (l_{Pl}/l_{\rm str})^2$ and the bound (\ref{XX}) becomes
\begin{eqnarray} \label{XXX}
\gamma^2_{\rm str}/a^2 \leq 3.904 \times 10^{11}(l_{Pl}/l_{\rm str})^2.
\end{eqnarray}
Alternatively, we may view this in terms of an effective (warped) string scale $\tilde{l}_{\rm str} = a^{-1} l_{\rm str}$. Hence, the current density required to produce black holes of mass $M = 2 \times 10^9 M_{\odot}$ by $z = 7.085$ depends on the ratio of the warped string scale $a^{-1}l_{\rm str}$ to the Planck scale in the wound string scenario.

Finally, we note that the subtle differences between the estimates of the upper bound on $\gamma^2$ obtained in Sec. \ref{Sect4} and those obtained here result from the incorporation of a model-dependent fine structure constant $\alpha_q = q^2/q_{Pl}^2$ in the expression for the string width (i.e., $r_c \approx (\sqrt{\alpha_q}\tilde{\eta})^{-1}l_P$) in the field-theoretic case. By contrast, in the wound string case, we have implicitly assumed that the compactification radius $R$ is independent of the elementary charge $q$ associated with higher-dimensional flux. However, incorporating a warp factor $a^2$, we note that, at least in the toy geometry \cite{Klebanov:2000hb}, this is related to $R^2$ via $a^2R^2 \sim \epsilon^{4/3}\alpha'$, where $\epsilon^{4/3}$ is the deformation parameter of the conifold in the extra-dimensional space \cite{Klebanov:2000hb,Lake:2010qsa}. Thus, by identifying $\tilde{\eta}^2 = (\tilde{m}_{\rm str}/m_{Pl})^2 = (l_{Pl}/\tilde{l}_{\rm str})^2$, where $\tilde{m}_{\rm str}$ and $\tilde{l}_{\rm str}$ denote the {\it warped} string mass and length scales, we obtain $\alpha_q \sim \epsilon^{-4/3}$, giving $R \sim \tilde{l}_{\rm str}/\sqrt{\alpha_q} \sim (\sqrt{\alpha_q}\tilde{\eta})^{-1}l_{Pl} \sim r_c$. Under these identifications, the chiral string model \cite{Witten85} and the dimensionally reduced wound string model in the Klebanov-Strassler geometry \cite{WS_KS} are exactly equivalent from a four-dimensional perspective. Nonetheless, even if, in general, subtle phenomenological differences between wound strings and chiral strings exist, it is clear that wound strings can also carry sufficient current to enable energy transfer from the string network to form the main accretion mechanism for early epoch SMBH.

\subsubsection{Refined analysis for higher-dimensional $F-$strings} \label{Sect5.2.2}

Accounting for the expansion of the black hole event horizon (as before) and imposing the limit $\lambda_z^{\rm min} = 2\pi R$ gives a correction to our previous estimates of order $\mathcal{O}(\tilde{\mathcal{T}}G/c^4) \sim (l_P/\tilde{l}_{\rm str})^2$, in the general case including a warp factor $a^2 \in (0,1]$. By Eq. (\ref{XXX}), this implies $\tilde{\mathcal{T}}G/c^4 \gtrsim 10^{-12}$, so that we may assume this contribution is negligible, except in pathological cases.

\section{How do seed black holes connect with the string network?} \label{Sect6}

String$-$black hole systems can form in two qualitatively different ways. First, the seed black holes and superconducting string network may form separately, before random collisions bring them into contact. In this scenario, either the strings or the primordial black holes (PBH) may form first, or there may be a continuous process of PBH formation $-$ for example, from primordial density fluctuations \cite{PBH} $-$ over a certain period of time. The string$-$black hole connection probability will therefore depend on several factors, including both the string and PBH number densities, their dynamics and decay rates, etc. However, given the extremely small physical size of PBH in most standard formation scenarios (e.g. for the initial mass of a PBH expiring at the present epoch, $M \approx 10^{14}$ g, $r_S \approx 10^{-16}$ m) and the extremely narrow width of strings formed at high symmetry breaking energy scales (e.g. for $m \gtrsim m_{GUT} \approx 10^{16}$ GeV/c${}^2$, $r_c \lesssim 10^{-32}$ m $\approx 10^3 l_{Pl}$) incredibly high number densities of both would be required in the early Universe, in order for a substantial number of collisions to occur.  We may therefore assume, without the aid of detailed calculations, that this process is at best subdominant, if it occurs at all.

Second, they may form ``together"; that is, already connected as a single system, though this scenario does not necessarily require the strings and seed black holes to form at exactly the same epoch. For example, it is possible for charged matter to become trapped in the string core during the process of network formation via the usual Kibble mechanism \cite{VS_book}. Hence, if a charged black hole such as that considered in \cite{Ary86} were to exist in the presence of a locally charged ISM, it is possible for a black hole$-$superconducting string system to form at the phase transition epoch. That said, we note that, at present, no explicit solution to Einstein's field equations representing such a system exists, though its construction would undoubtedly represent a worthwhile contribution to the literature. Nonetheless, since there do not seem to be any fundamental physical barriers to the existence of such a solution, we may assume this as a viable mechanism for the production of a seed black hole$-$superconducting string system. 

The properties of Abelian-Higgs strings, in which a small amount of ``external" charged fluid becomes trapped within the string core, were studied by Lake and Yokoyama \cite{WS_KS}. The basic ansatz is of the form
\begin{eqnarray} \label{ansatz*}
\phi(r,\theta,z,t) &=& \eta f(r) \exp \left[in\theta + i\psi(z,t)\right], \ A_{\theta}(r) = \frac{n}{e}a_{\theta}(r)
\nonumber\\
A_{z}(r,z,t) &=& \frac{n}{e}a(r)\psi'(z,t),
\nonumber\\
A_{0}(r,z,t) &=& \frac{n}{e}a(r)\dot{\psi}(z,t),
\end{eqnarray}
where $f(r)$ and $a_{\theta}(r)$ are dimensionless functions obeying (\ref{NO_f_bc}) and the new function $a(r)$ also obeys analogous boundary conditions. In order to prevent divergences in the electromagnetic flux and energy densities as $r \rightarrow 0$, it is necessary to add a term proportional to $A^{\mu}J_{\mu}$ to the usual Abelian-Higgs Lagrangian, where $J^{\mu}$ represents a charged fluid. Setting
\begin{eqnarray} \label{ansatz**}
J_{0}(r,z,t) &=& \frac{n}{e}J(r)\psi'(z,t),
\nonumber\\
J_{0}(r,z,t) &=& \frac{n}{e}J(r)\dot{\psi}(z,t),
\end{eqnarray}
where $J(r)$ is nonzero only in a small region $r \in [0,\delta]$, $\delta \leq r_c$ (not to be confused with the angular deficit induced in the spacetime by the presence of the string), the EOM may be solved to leading order, as before, and the components of the energy-momentum tensor computed \cite{WS_KS}. It is straightforward to verify that the consistancy of the ansatz (\ref{ansatz*}) with the covariant EOM requires the function $\psi(z,t)$ to obey Eq. (\ref{HD_string_EOM}), so that the constant current case corresponds to Eq. (\ref{linear_psi}). This scenario was studied in detail in \cite{WS_KS}, where it was found that the physical parameters of the string take exactly the same form as those given in Eqs. (\ref{HD_E}) and (\ref{HD_Q}), but with $\mathcal{T}$ replaced by $\mu \sim 2\pi \eta^2|n|$, as in the standard Nielsen-Olesen solution. In this case, $\psi(z,t)$ plays exactly the same role as $\Gamma(z,t)$ in the chiral string model, except that the resulting winding number $m$ is not a topological invariant. Hence, the higher-dimensional analogue of the model presented in \cite{WS_KS} is the same as that given in Sec. \ref{Sect5}, where the strings wrap windings in a simply connected internal space.

However, even if no ``external" charged fluid becomes trapped in the string at the epoch of formation, it may still be possible for a black hole$-$superconducting string system to form spontaneously. For example, in the $U(1)\times \tilde{U}(1)$ model, the black hole in \cite{Ary86} may carry $U(1)$ charge, so that superconducting chiral strings emanating from its horizon may form at the $U(1)$ symmetry-breaking phase transition. We assume such a system is viable in general relativity, as there exists no fundamental physical impediment to its formation. Alternatively, it has been argued that quantum gravity effects impose a natural cut-off for the functions $f(r)$ and $a_{\theta}(r)$ in the Nielsen-Olesen ansatz, at some scale $r = \delta \sim l_{Pl}$, allowing local $U(1)$ strings to support a zero mode \cite{Svetovoy:1997dk}. If correct, this would allow Abelian-Higgs strings to become superconducting $-$ including those formed via the mechanism proposed in \cite{Ary86}. Spontaneous current generation in the string network, in the absence an external field, was also demonstrated by Peter, and may be capable of converting non-superconducting string--black hole systems, such as those studied in \cite{Ary86,refp}, into superconducting ones.


Finally, another possibility is that seed black holes form \emph{on} the string network, via some form of high energy interaction. This is especially likely in networks of cosmic ``necklaces" $-$ systems of monopoles connected by strings, which exist in both field-theoretic scenarios and models based on higher-dimensional string theory \cite{Lake:2010qsa,CS_NECKLACE} $-$ in which high-energy collisions between monopoles (or ``beads") can generate PBH. In addition, black hole pair-production may take place when a cosmic string snaps, though such models have yet to be generalized to the superconducting case \cite{string_PBH}.

It is beyond the scope of this paper to investigate the possible formation processes for black hole$-$superconducting string systems in detail. However, we note that several theoretically viable mechanisms exist, and these cannot be discounted \emph{a priori}. Nonetheless, both accretion from the ISM and mergers may still play a vital role in the evolution of SMBH from the seed black hole population, regardless of whether energy absorption from superconducting strings is the dominant accretion mechanism.

\section{Cosmological implications} \label{Sect7}

By assuming that $\epsilon _L$, $\epsilon _M$ are constants, Eq.~(\ref{1}), giving the time variation of the SMBH mass in the standard scenario, can be integrated to give the following exponential law for the mass increase:
\be
M(t) = M\left(t_i\right)\exp\left[C\left(t-t_i\right)\right],
\ee
where $M\left(t_i\right)$ is the black hole mass at $t=t_i$, $C=\epsilon _L\left(1-\epsilon _M\right)/\epsilon _M\tau $, and $\tau \approx 1.45\mu _e^{-1}\times 10^{16}$ s \cite{Shap}. By assuming that the most common merger events are between two black holes of comparable mass, the mass after $\mathcal{N}$ merging events is $M_{\mathcal{N}}(t) = M\left(t_i\right) f_1...f_{\mathcal{N}}\exp\left[C\left(t_f - t_i\right)\right] \equiv f \exp\left[C\left(t_f - t_i\right)\right]$, where $f_{\mathcal{N}} = M\left(t_{\mathcal{N}}\right)/M_{{\mathcal{N}}-1}\left(t_{\mathcal{N}}\right)>1$ is the mass amplification factor and $t_f$ is the time at which black hole growth ends.

By starting with an initial SMBH seed mass of order $M\left(t_i\right)=5M_{\odot}$ and assuming that the luminosity of the quasar is at the Eddington limit $\epsilon _L=1$, the mean radiative efficiency of the accretion onto black holes is $\epsilon _M=0.3$ \cite{Shap}. By taking the value $\mu _e=1/(1-Y/2)=1.14$ for the mean molecular weight, the mass of the black hole after a time interval $\Delta t = t_f-t_i=1\;{\rm Gyr}=3.1\times 10^{16}$ s is then $M=8.93\times 10^2M_{\odot}$. This increases to $M=3.625\times 10^4M_{\odot}$ for $\epsilon _M=0.2$. Therefore, in the standard scenario for SMBH formation, either super Eddington accretion, $\epsilon _L \gg 1$, or huge mass amplification factors of order  $f\approx 10^5-10^7$ are required. By contrast, as one can see from Eq.~(\ref{bound}), in the model of a seed black hole pierced by a two superconducting strings, the black hole can acquire a mass of order $10^{9}M_{\odot}$ in a time interval of order $10^{16}$ s, if the string parameters that control the current and intrinsic tension satisfy the constraint $(\gamma/\kappa)^2 \lesssim 10^{10}$. 

Unfortunately, the number of SMBH discovered at redshifts $z\geq 6$ is still rather limited, being at present around $40$ \cite{Q1}. Hence, it is not easy to infer their true number density in the early Universe using the current observational data. An estimate of the comoving number density of the population at $z \geq 6$ was given in \cite{est} as $n_{\rm SMBH}=1.44\times 10^{-5}$ Mpc$^{-3}$. By assuming a mean SMBH mass of $M_{\rm BH} \approx 10^{10 }M_{\odot}$, it follows that the corresponding mass density of black hole matter can be estimated at redshift $z\approx 6$ as $\rho _{\rm SMBH}\approx 1.44\times 10^5M_{\odot}$ Mpc$^{-3}$. Therefore, if we assume energy transfer from superconducting strings to be the dominant accretion mechanism, this figure also represents the energy density extracted from the string network.

The absolute number of SMBH within the horizon, at time $t$, is given approximately by $N_{\rm SMBH} \approx n_{\rm SMBH} \times (ct)^3$, so that $N_{\rm SMBH} \approx 200$ at $z \approx 6$. Next, we use the fact that $M_{\rm BH} \approx j^2 \mu L_0$, where $L_0$ is the total length of string over which current must flow into a single seed, in order to form an SMBH of mass $M_{\rm BH}  \approx 10^{10}M_{\odot}$, to obtain $L_0 \approx  9.187 \times 10^{47} \sqrt{\hbar c/G} \times j^{-2}\mu^{-1}$. The total string length from which current is absorbed, in order to create all the SMBH present at $z \approx 6$, is therefore
\bea
L_{\rm tot} \approx 7.065 \times 10^{56} \sqrt{\frac{\hbar G}{c^3}} \times j^{-2}\left(\frac{\mu}{\mu_{\rm CMB}}\right)^{-1},
\eea
where we have normalized the string tension using the standard upper bound $G\mu_{\rm CMB}/c^2 = 2.6 \times 10^{-7}$, obtained from the fitting of (non-superconducting) vacuum string models to the CMB and SDSS data \cite{string_tens_bounds}. For $j^2 \approx 1$, this implies a total length $L_{\rm tot} \approx 4.431 \times 10^{21} \rm m \sim 10^5 \ pc$ for GUT scale strings, or, equivalently, $L_0 \sim 10^3$ pc per seed.

We may also estimate the number of string loops present at this time. Numerical simulations and analytic models of cosmic string evolution show that string networks evolve in a self-similar manner \cite{cs}. It follows that, at time $t$, a horizon-sized volume of the Universe contains only a few long strings, stretching across the entire observable Universe, together with a large number of small closed string loops. These loops,  having recently ``chopped off" from the string network, may then decay via gravitational radiation or gauge particle emission \cite{Tye}. The number density of the loops is given by \cite{Vilc}
\be \label{loop_dens}
n_{l}(t) \approx \zeta \alpha _s^{-1}(ct)^{-3},
\ee
where $\xi(t) = \zeta^{-1/2}ct$ is the correlation length of the string network, the average loop size is given by $L = \alpha_s ct$, where $\alpha _s\geq k_gG\mu/c^2$, and $k_g$ is a numerical factor of order $k_g \approx 50$, determined by simulations \cite{cs,Vilc}. The exact values of the parameters $\zeta$ and $\alpha _s$ are not known, but current numerical simulations of vacuum strings suggest $\zeta \sim \mathcal{O}(10)$ and provide an upper limit of $\alpha_s \lesssim 5 \times 10^{-3}$, though it is not clear whether these extend to the superconducting case.

Assuming that Eq. (\ref{loop_dens}) holds, at least approximately, for superconducting strings, the average mass contained in an individual loop is
\begin{eqnarray}
M_l(t) \approx (1+j^2) \mu \alpha_s ct,
\end{eqnarray}
so that the mass density associated with all the string loops at time $t$ is
\bea
\rho _l(t) &=& M_l(t)n_l(t) \approx (1+j^2) \mu (ct)^{-2}
\nonumber\\
&=& 2.6 \times 10^{-6} \left(\frac{G}{c^2}\right)^{-1} (1+j^2) \left(\frac{\mu}{\mu_{CMB}}\right)(ct)^{-2},
\nonumber
\eea
\bea
{}
\eea
where we have set $\zeta = 10$ for simplicity. For vacuum strings, the energy per unit length $\mu c^2$ and the absolute value of the string tension $|\mathcal{T}|$ are equal, but for the superconducting strings in our model we have $\mu c^2 = \mathcal{T}(1+\gamma^{-2})$ and $|\mathcal{T}^{z}| = \mathcal{T}(1-\gamma^{-2})$. However, the bound obtained from CMB+SDSS fitting also applies to the \emph{intrinsic} tension of ``wiggly" strings \cite{string_tens_bounds}. These obey a similar equation of state to superconducting strings, in which the effective mass per unit length is increased while the effective tension is decreased \cite{small-scale-structure}. Therefore, there is good reason to believe that it holds for the intrinsic tension of superconducting strings also, at least to within an order of magnitude. At $t \approx10^{16}$ s, the total mass contained in the string loops is, therefore,
\begin{eqnarray} \label{Mtot}
M_l^{\rm tot}(10^{16} {\rm s}) \approx 5.249 \times 10^{15} (1+j^2) \left(\frac{\mu}{\mu_{\rm CMB}}\right) \ M_{\odot}.
\end{eqnarray}

Next, we modify the standard string network evolution picture by assuming that some fraction of the total mass contained in the string loops, at each epoch within the range $0 \lesssim t \lesssim 10^{16}$ s, is accreted to the seed black hole population. Since the mass contained in the loops, at any time $t$, must be less than the total mass of the network at the epoch of string formation ($t_s \approx 0$), the resulting estimates may be considered as lower bounds on the true accretion rate.

Thus, at $z \approx 6$, the remaining mass contained in the loops may be modelled as $\mathcal{M}_{l}^{\rm tot} = (1-\epsilon_l) M_{l}^{\rm tot}$, where $0 < \epsilon_l \leq 1$ is the fraction of $M_{l}^{\rm tot}$, given by Eq. (\ref{Mtot}), absorbed by the SMBH. This is true generically, regardless of whether the accretion occurs as a result of current flow or by the absorption of string rest mass, though it is reasonable to assume current flow to be the dominant mechanism. In this case, we may equate $\epsilon_l M_{l}^{\rm tot}$ with the total mass present in the SMBH at $t \approx 10^{16}$ s, $M_{\rm BH}^{\rm tot} = j^2\mu L_{\rm tot}$, giving
\begin{eqnarray}
\epsilon_l \mu \approx 3.5 \times 10^{35} (1+j^2)^{-2}\left(\frac{M_{\odot}}{m_{Pl}}\right)^{-1}\mu_{\rm CMB}.
\end{eqnarray}
For $\epsilon_l \rightarrow j^2$, the limit in which the seed black hole population absorbs {\it all} the current within the loops, we have $G\mu/c^2 \approx 2.441 \times 10^{-10} j^{-2}$, so that $\mu \approx \mu_{\rm CMB}$ requires $j^2 \approx 9.388 \times 10^{-4}$. By contrast, for $\epsilon_l \rightarrow 1$, the limit in which the black holes absorb the loops themselves, we require only $G\mu/c^2 \sim 10^{-10}$, irrespective of the value of $j^2$.

However, in the context of our idealized model, there is a (potential) problem in assuming that the black hole gains mass by absorbing string loops: single loops piercing the black hole cannot deliver equal but opposite charge to different points on the horizon, since this would violate charge conservation. As before, we need at least two strings, with zero net charge and momentum. Nonetheless, our model approximates mass-energy accretion by black holes swallowing loops if the initial curvature radius of each loop is large compared to the Schwarzschild radius. Antipodal strings piercing the black hole are then seen as idealized approximations representing accretion from large loops and, since both the net charge and momentum of the whole loop population are expected to be zero, and SMBH population with zero net charge and momentum may also be formed.

Interestingly, if we set the loop formation parameters equal to their threshold values, $\zeta \sim 10$, $\alpha_s \sim 10^{-3}$, this implies the existence of $N_{l} \sim 10^4$ loops at $t \sim 10^{16}$ s, each of length $l \sim 3 \times 10^{23}$ cm $= 0.1$ Mpc, giving a total length $l_{\rm tot} \sim 3 \times 10^{27}$ cm $= 10^3$ Mpc contained in the loops. The average length of a long string is then of order $\xi \sim 0.1 \times ct$, so that $\sim 10^2$ long strings, each of length $\xi \sim 10$ Mpc, contain the same total length (and mass) as the loops. In this case, setting $\mu \sim \mu_{\rm CMB} \sim \mu_{\rm GUT}$ and $\epsilon_l \sim j^2 \sim 10^{-3}$, a population of $N_{\rm BH} \sim 10^2$ seed black holes need be {\it initially} ``plugged in" to only $\sim 10^2$ long strings, each of which grows in proportion to the horizon size, maintaining a constant of proportionality $\sim 0.1$, unless it is completely fragmented by loop production. However, even if {\it all} such strings fragment completely into $N_{l} \sim 10^4$ loops, by $z \sim 7$, this process can still account for approximately $1\%$ of SMBH mass accretion. With only a modest increase in the value of $\epsilon_{l}$, for example $\epsilon_l \rightarrow 10^{-2}$, string-fuelled accretion can still form the dominant contribution to SMBH formation, even under such stringent restrictions.

On the other hand, were we to imagine a black hole pierced by two genuinely ``long" (i.e. cosmological horizon-crossing) strings, we may, at late times, encounter a very real physical problem for our mass accretion model. In this case, there is no mechanism by which the black hole growth can end - at least, not until the ``end point" of the string crosses the cosmological horizon and is subsequently swallowed by the event horizon of the black hole.

At present, neither the loop accretion nor long string accretion scenarios can be ruled out by available data, though further work, including numerical simulations of string accretion models, is needed in order to determine which is more plausible. In either case, since the mass-energy contained in the loops at any time $t$ must be less than the mass-energy contained in the network at the epoch of string formation, the calculations above yield a lower bound for the total mass transfer from the string network to the seed black hole population.

\section{Astrophysical realizations of the model}\label{Sect8}

In this section, we consider a number of important issues which are crucial to the implementation of our model in {\it any} realistic astrophysical/cosmological setting. In particular, we pose, and propose tentative solutions to, a number of questions regarding the specific physical conditions and astrophysical processe(s) that must be realised in order for a sufficient number of superconducting string--black hole systems to form in the early Universe, and to persist for time periods long enough to lead to sufficient mass accretion. In so doing, we clarify a number of subtle conceptual issues inherent in the proposed mechanism.

For example, it may be argued that since, in ``standard" scenarios, current-carrying strings (and even topological defect strings {\it without} current) cannot have end points, it is impossible for them to ``end" on the horizon of a black hole. However, in our model, the strings ``end" on the event horizon only from the perspective of an observer located outside the black hole. From the perspective of the string itself $-$ or, more specifically, of an observer comoving with a fluid element travelling through the string $-$ nothing unusual happens at the horizon, with increased tidal forces being the only locally observable effect. Thus, all conservation laws, including conservation of the topological winding number for defect strings, are satisfied. Physically speaking, the strings ``end" only on the singularity inside the black hole, but this does not contradict any topological charge, electric charge, and/or momentum conservation requirements for either defects or $F-$strings.

Nonetheless, in a realistic setting, our model must confront the fact that strings carrying oppositely charged currents will, if they intersect on (or near) the surface of the black hole, unzip and annihilate \cite{string_unzip}. {\it This is a viable astrophysical process which must be considered in any physically realistic model.} As such, it is necessary to consider two important sub-questions: (i) What is the probability that two strings, piercing the black hole horizon, will intersect within a given time interval $t$? (ii) If the strings do intersect, how long will it take for them to unzip an annihilate? (In other words, for how long, prior to total annihilation, will they be able to continue supplying current and mass to the black hole?)

Assuming that the string ``end points" undergo a random walk on the surface of the horizon with root mean square velocity $\sqrt{\langle v^2 \rangle} \leq c$, the maximum step length within the characteristic time $\tau = r_c/c$ is simply $r_c$, the string width. The total displacement from the initial position in a time interval $\Delta t = t-t_s$, where $t_s$ is the epoch of formation for the string--black hole system, is then
\begin{eqnarray} 
d(t-t_s) \approx \sqrt{r_c c(t-t_s)} + r_c \equiv d(N) \approx r_c\left(\sqrt{N} +1\right) \,
\nonumber
\end{eqnarray}
\begin{eqnarray} \label{random_walk}
{}
\end{eqnarray}
where $c(t-t_s) = N\tau$ and $N$ is the number of steps. The additive term $r_c$ on the right-hand side of Eq.  (\ref{random_walk}) accounts for the intrinsic width of the string, which, if greater than the initial radius of the black hole (at $t=t_s$), makes collision inevitable. Similarly, at time $t>t_s$, the radius of the black hole is
\begin{eqnarray} \label{PBH_rad}
r_S(t-t_s) \approx r_S(t_s) + j^2\mu (G/c^2) c(t-t_s)  \, ,
\end{eqnarray}
where $r_S(t_s)$ is the radius at the string formation epoch (In performing this estimate, we are no longer free to assume a negligible initial mass for the black hole, which must now be accounted for explicitly.) Thus, assuming an average initial distance between the end points of order $r_S(t_s)$ (which is reasonable for randomly placed strings, and which would result in a separation of $r_S(t-t_s)$, for $t>t_s$, if neither point were to move relative to its initial position) a necessary, {\it but not sufficient}, condition for the pair to intersect is that $|r_S(t-t_s) - d(t-t_s)| \lesssim r_c$, giving
\begin{eqnarray} \label{cond-1}
j^2\mu (G/c^2) c(t-t_s)  \lesssim  \sqrt{r_c c(t-t_s)} + 2r_c - r_S(t_s)\, .
\end{eqnarray}
Thus, if $r_S(t_s) < 2r_c$, intersection is inevitable at $t=t_s$.

For the sake of both simplicity and concreteness, we here neglect more complex and/or speculative formation mechanisms for the string--black hole system, and, using the standard model of PBH formation from collapsing over-densities in the early Universe, we have \cite{PBH}
\begin{eqnarray} \label{M_PBH}
M_{\rm PBH}(t) \approx \frac{c^3}{G}t \, ; \quad  r_{S}(t) \approx 2ct \, .
\end{eqnarray}
Thus,
\begin{eqnarray} \label{}
r_{S}(t_s) \approx 2r_c \, ,
\end{eqnarray}
and the condition (\ref{cond-1}) reduces to
\begin{eqnarray} \label{cond-2}
t -  t_s \lesssim  \frac{r_c c^3}{j^4\mu^2 G^2} \approx \frac{(2\pi)^4r_c^5}{l_{Pl}^4c} \times \alpha_q^2\left(\frac{\gamma}{\kappa\beta}\right)^4 \, ,
\end{eqnarray}
where the final equality applies to the defect string model considered in Sec. \ref{Sect4.1}-\ref{Sect4.2}. Using the bound $(\gamma/\kappa)^2 \leq 6.213 \times 10^{10}\alpha_q^{-1}$ (\ref{bound}) then gives
\begin{eqnarray} \label{cond-2}
t -  t_s \lesssim 6.016 \times 10^{24} \tau \, .
\end{eqnarray} 

For GUT scale strings, $\tau = r_c/c \approx 10^{-31}$ m, which corresponds to an initial black hole mass of order $m_{Pl}^2/m_{\rm GUT} \approx 0.1$g, so that intersection of the string end points is possible only over a time period of $\Delta t = t-t_s \lesssim 10^{-7}$ s from the time of formation of the string--black hole system. We may place a crude lower bound on the {\it maximum} probability of intersection within this interval by dividing the total area traced out by the string cross section, on the surface of the horizon within $\Delta t$, by the average value of the horizon area within the same period:
\begin{eqnarray} \label{Prob}
P_{\rm max} \sim \frac{r_cc\Delta t}{(r_c+c\Delta t)^2} \sim \frac{r_c}{c\Delta t} \gtrsim 10^{-25} \, .
\end{eqnarray}
However, in reality, the condition (\ref{cond-1}), from which (\ref{Prob}) is derived, is a necessary condition only for the {\it paths} of the two end points to intersect. The actual intersection of the end points requires the paths to intersect at the same moment in time, which is a much stronger condition \cite{Lawler}.

Since we have assumed that the string are ``long" (i.e., that their curvature radius is far larger than the initial radius of the black hole), it follows that the the minimum distance between them falls of as $r^{-1}$, where $r$ is the radial coordinate measured outward from the centre of the black hole. Hence, the probability of collisions between strings is greatest, but still extremely small, at the horizon, and quickly vanishes due to the fact that the horizon size grows proportional to $t$, whereas the random walk induces an end point displacement proportional to $\sqrt{t}$.

Nonetheless, we may consider the situation in which intersection does occur and ask: How much mass can the superconducting strings still transfer to the black hole before total annihilation occurs? Assuming an arbitrary unzipping speed, $v_{\rm unzip} \leq c$, a pair of ``long" strings with characteristic length $L$ will annihilate completely within a time scale $\Delta t \sim L/v_{\rm unzip}$. However, the length of the strings is set by the correlation length of the network, which is itself a function of time, according to $\xi(t) = \zeta^{-1/2}ct$, where $\zeta \sim \mathcal{O}(10)$. Thus, in order for the unzipping process to ``outrun" the expansion of the correlation length of the network, we require $v_{\rm unzip} \geq \zeta^{-1/2}c$. In this case, the unzipping will be complete within a time interval
\begin{eqnarray} \label{}
\Delta t \approx \frac{\zeta^{-1/2}c t_s}{v_{\rm unzip} - \zeta^{-1/2}c}\, .
\end{eqnarray}
For $v_{\rm unzip} \gg \zeta^{-1/2}c$ this interval is even shorter than $t_s$, the formation time of the string network, but, as $v_{\rm unzip} \rightarrow \zeta^{-1/2}c$, it can become very large, potentially leaving enough time for the strings to transfer a significant fraction of their mass-energy to the seed black hole. Though further investigation of this scenario, which would require detailed estimates of $v_{\rm unzip}$ provided by simulations, is beyond the scope of this paper, we note that $\Delta t \gtrsim 10^{16}$ s requires $v_{\rm unzip} \lesssim \zeta^{-1/2}c(1 + t_s/10^{16} {\rm s})$. In other words, in order for the unzipping process to take longer than 900 million years, $v_{\rm unzip}$ must be within $\sim 10^{-38}\%$ of its critical value for GUT scale strings, so that extreme fine tuning is required to realize this possibility.

In principle, these conclusions may be altered by the presence of small-scale structure on the strings \cite{VS_book}, so that the probability of string intersection remains non-negligible within a region $r \lesssim l_{\rm k}$, where $l_{\rm k}$ is the average length of a small-scale ``kink" on the string. However, in general, it is clear that the linear growth of the event horizon will quickly ``outrun" the displacement due to the random walk of the string end point, and we may conjecture that the growth of kinks is unlikely to counter this effect, due to a combination of the stretching of the network with the Hubble flow \cite{VS_book}, together with the ``smoothing" effects of the current.

\section{Discussion and final remarks}\label{Sect9}

In the present paper we have considered an alternative scenario for SMBH formation in the early Universe, at redshifts $z\geq 6$. The standard $\Lambda$CDM cosmological model faces severe difficulties in explaining the presence of supermassive objects with masses of order $10^{10}M_{\odot}$ at such early times. In particular, the Millennium Simulation, based on standard $\Lambda$CDM cosmology, predicts a mass density for SMBH of order $10^{-9} M_{\odot} \rm \ Mpc^{-3}$, large enough to give only $\sim \mathcal{O}(1)$ SMBH per horizon volume at redshifts $z \gtrsim 6$ \cite{est}.

We have proposed that superconducting cosmic strings, which may have been produced in large numbers at symmetry breaking phase transitions in the post-inflationary era, play an essential role in SMBH formation. Strings and black holes may form complex systems, introduced for the first time in \cite{Ary86}, in which long strings pierce the black hole, forming a stable general relativistic system. In the important case of the $U(1)$ Abelian-Higgs model, where only numerical solutions of the field equations can be obtained, this string$-$black hole system may be interpreted physically as representing a black hole with long-range string ``hair". This ``hair" can act as an energy source for the initial seed black hole, thus allowing a rapid but steady growth in rest mass.

Adopting a simple phenomenological model of the superconducting string, we have estimated the energy transport to the black hole, via an electric current, and have obtained an estimate for the mass increase. Even though the time variation of the mass is linear, a very rapid increase can be achieved due to the high rate of energy transfer from the cosmic string, if the current density is sufficiently high. As compared to the standard accretion-merger model for SMBH growth, the string-pierced black hole model naturally allows the formation of SMBH at epochs as early as 0.77 Gyr, assuming that the composite systems form in the very early Universe, and survives over this time period.

The results obtained depend on the precise values of three dimensionless string model parameters $\left(\kappa , \beta, \gamma \right)$, that are related physically to the string tension, width and current density, respectively, which together control the energy transfer to the black hole. Conversely, the existence of SMBH in the early Universe allows us to test the parameters of the string model, thus imposing a distinct (and new) set of constraints on the fundamental structure of the underlying field theory or string theory. 

From a cosmological point of view, we may assume that a large fraction of the mass of the strings formed during the inflationary/post-inflationary era ``disappeared" inside the SMBH through energy transfer to the primordial seed black holes. Such an effect, together with the gravitational or electromagnetic decay of string loops, may have significantly contributed to the drastic decrease in the number of strings in the present day Universe.

The assumption that the string--black hole systems forms at, or soon after, the epoch of string formation, and survive to a redshift of $z \sim 6-7$, may reasonably be regarded as the weakest point of our analysis. However, we note that, from a phenomenological perspective, this assumption was made in order to allow us to perform concrete calculations, and is in no way fundamental. Specifically, assuming the formation of the string--black hole system at $t_s \sim 0$ gives a definite time interval of $\sim 900$ million years, which then allows the minimum necessary current that must delivered to the black hole (over that period) to be determined.

Clearly, this this assumption is not a necessary requirement of our proposed mechanism, and we could just as well have assumed a limiting current -- for example, the threshold current determined by Witten \cite{Witten85}, or the value suggested by the motion of the string network in the primordial magnetic field (c.f. Sec. II) -- and used this to determine the minimum time period over which the string--black hole system must survive intact. As noted in Sec.~\ref{Sect2.1}, for GUT-scale strings carrying the threshold current, the string--black hole system need only survive for $\sim 1000$ years. This allows our initial the assumption about the formation of the string--black hole system at, or close to, the epoch of string formation to be relaxed by many orders of magnitude. In our proposed mass-transfer mechanism, higher currents require shorter time-intervals and vice versa.

In addition, we note that there are several viable scenarios able to realize the formation of a string--black hole system. These include the formation of: (a) chiral strings, emanating from an initially charged black hole, (b) Abelian-Higgs strings, emanating from an initially charged (or uncharged) black hole, in which charged fluid becomes trapped in the string core during the process of string formation via the usual Kibble mechanism, (c) hypothetical ÒregularizedÓ Abelian-Higgs strings, capable of supporting a zero mode, as suggested in \cite{Svetovoy:1997dk}, (d) string--black hole systems formed via the collision of monopoles/``beads" on cosmic necklaces \cite{Lake:2010qsa,CS_NECKLACE} and (e) string--black hole systems formed by the ÒsnappingÓ of a string \cite{string_PBH}. Current may also be generated spontaneously within the string network, even in the absence of an external field \cite{Peter:1993tm}, converting non-superconducting string--black hole systems into superconducting ones.

In obtaining the present order of magnitude estimates for mass transfer from the string network to the SMBH, we have neglected the possible effects of the black hole's gravitational field on the energy transfer processes, as well as the gravitational field of the strings. Nonetheless, such effects could certainly play an important role in the energy transfer processes. Furthermore, standard accretion and merging processes may also have played an important role in the growth of the seed black holes. 

It is also important to remember that the strings are part of a network whose dynamics affects the lifetime(s) of the superconducting string sections connected to the seed black holes. We addressed this by considering the rate at which the network fragments into loops, which subsequently decay via gravitational and/or electromagnetic radiation. Since loop sizes must be comparable to, or less than, the correlation length of the network (which is time-dependent), our model requires loop lifetimes to be long enough for significant mass transfer to the string--pierced black hole to occur. However, no attempt was made to model the entire network as string--black hole system. Nonetheless, it must be noted that, if black holes do intersect with the network and grow linearly in time, as we have proposed, this in itself may significantly affect the network evolution.

Thus, in order to truly test the viability of our proposed mechanism, it is necessary to construct analytical and/or numerical simulations of black hole accretion and mergers, for black holes connected to a superconducting string network. A unified black hole growth process, including energy transfer from strings, accretion and merging, could give a more realistic astrophysical description of the black hole mass dynamics during the early stages of the cosmological evolution.

Such a model may potentially solve, or at least alleviate, two problems at once: the lack of direct observational evidence for strings, in the present day or at high redshifts, together with the low mass density of SMBH predicted by $\Lambda$CDM. The latter poses a severe observational challenge to the standard model of cosmology, while the former is a puzzle for high energy theorists, given the ubiquitous production of strings predicted by many GUT theories \cite{Jeannerot:2003qv}. It may therefore be hoped that a combined $\Lambda$CDM + strings model could offer a better explanation of the available astrophysical and cosmological data. With this in mind, we note it has already been shown that a $\Lambda$CDM Universe containing strings offers a marginally better fit to the available CMB power spectrum \cite{CS_CMB}.

Finally, we note that other types of string-like objects, not limited to topological defects and/or $F$- and $D$-strings predicted by string theory, may have been present in the early Universe. In \cite{Harko:2014pba}, it was suggested that Bose-Einstein condensates (BECs), consisting of axionic dark matter or cosmological scalar fields \cite{Davidson:2014hfa}, may have condensed to form string-like objects on cosmic scales. Such objects are capable of fulfilling the role of dark matter filaments \cite{DM_filaments}. In this context, it is significant that two-component BECs are capable of forming ``twisted" string states, analogous to superconducting chiral strings, or higher-dimensional wound-strings considered in the present work \cite{BECvortons}. BEC string analogues of cosmic vortons have even been observed in the laboratory \cite{BECvortons} and their existence on cosmological scales and/or in the early Universe, cannot be discounted {\it a priori}. The investigation of momentum/mass-energy transfer to seed black holes from dark matter filaments is therefore of great theoretical interest.

In addition, it must be noted that the results presented above may be significantly altered in modified gravity theories, many of which have been proposed as alternatives to the $\Lambda$CDM concordance model (see, for example, \cite{mod_grav} and references therein), since any modification of canonical general relativity may significantly affect the gravitational properties of both black holes \cite{BH_mod_grav} and strings \cite{string_mod_grav}. Theoretical investigations of black hole growth, string network evolution, and superconducting string--black hole systems in generalized gravity models may therefore be fruitful avenues for future research.

\section*{Acknowledgments}

We thank the anonymous referee for their insightful comments, which helped us to greatly improve the manuscript. TH thanks the Institute for Fundamental Study for the kind hospitality offered during the preparation of this work and ML thanks the Yat Sen School and the Department of Physics at Sun Yat Sen University for gracious hospitality during the final preparation of the manuscript. ML is supported by a Naresuan University Research Fund individual research grant.


\end{document}